\documentclass[11pt]{article}
\pdfoutput=1
\usepackage{jheppub}
\usepackage{latexsym}
\usepackage{amsmath,amsfonts,amsbsy,amssymb,amscd}
\usepackage{slashed}
\usepackage{graphicx,epsfig,picture,graphics}
\usepackage{float}                          
\usepackage{lscape}                         
\usepackage{bm}
\usepackage{tikz}

\usepackage[T1]{fontenc} 
\usepackage[utf8]{inputenc} 
\usepackage{microtype} 
\usepackage{lmodern} 
\usepackage{booktabs} 
\usepackage{mdframed} 
\usepackage{todonotes} 

\usepackage{tensor} 
\usepackage{braket} 
\usepackage{mathrsfs} 

\usepackage{stmaryrd}
\usepackage{enumerate}

\usepackage{amsthm} 
\newtheorem{theorem}{Theorem}

\def\comma{\,,\,}

\providecommand*{\dd}{\mathop{}\!d}
\renewcommand*{\dd}{\mathop{}\!d}

\providecommand*{\R}{{\mathbb{R}}}
\renewcommand*{\R}{{\mathbb{R}}}

\newcommand\parmp{\mathbin{\vcenter{\hbox{%
  \oalign{$\scriptstyle({-})$\cr
          \noalign{\kern-0.8ex}
          \hfil$\scriptscriptstyle+$\hfil\cr}%
}}}}

\providecommand{\Gt}{{\tt G}}
\renewcommand{\Gt}{{\tt G}}
\providecommand{\Jt}{{\tt J}}
\renewcommand{\Jt}{{\tt J}}
\providecommand{\Ht}{{\tt H}}
\renewcommand{\Ht}{{\tt H}}

\providecommand{\Pt}{{\tt{P}}}
\renewcommand{\Pt}{{\tt{P}}}

\usepackage{pifont}

\usepackage{tikz}
\usepackage{tikz-3dplot}

\newcommand*{\shi}{1}
\newcommand*{\shih}{0.5}
\newcommand*{\shihh}{0.25}

\newcommand{\eq}[2]{\begin{equation} #1 \label{#2} \end{equation}}
\newcommand{\vp}{\varphi}
\newcommand{\eps}{\epsilon}

\newcommand{\extd}{\dd}

\newcommand{\hJ}{\hat{{\tt J}}}
\newcommand{\hP}{\hat{{\tt P}}}

\title{Three-dimensional Spin-3 Theories  Based on General Kinematical Algebras}

\author[a]{Eric Bergshoeff,}
\author[b]{Daniel Grumiller,}
\author[b]{Stefan Prohazka}
\author[c,d]{and Jan Rosseel}

\affiliation[a]{Van Swinderen Institute for Particle Physics and Gravity,
University of Groningen,
\\Nijenborgh 4, 9747 AG Groningen, The Netherlands}
\affiliation[b]{Institute for Theoretical Physics, TU Wien,
\\ Wiedner Hauptstrasse 8--10/136, A-1040 Vienna, Austria}
\affiliation[c]{Albert Einstein Center for Fundamental Physics,
University of Bern,
\\Sidlerstrasse 5, 3012 Bern, Switzerland}
\affiliation[d]{Faculty of Physics, University of Vienna,
\\Boltzmanngasse 5, A-1090 Vienna, Austria}

\emailAdd{e.a.bergshoeff@rug.nl}
\emailAdd{grumil@hep.itp.tuwien.ac.at}
\emailAdd{prohazka@hep.itp.tuwien.ac.at}
\emailAdd{rosseelj@gmail.com}

\preprint{TUW--16--26}

\abstract{%
We initiate the study of non- and ultra-relativistic higher spin theories. For sake of simplicity we focus on the spin-3 case in three dimensions. We classify all kinematical algebras that can be obtained by all possible In\"on\"u--Wigner contraction procedures of the kinematical algebra of spin-3 theory in three dimensional (anti-) de~Sitter space-time. We demonstrate how to construct associated actions of Chern--Simons type, directly in the ultra-relativistic case and by suitable algebraic extensions in the non-relativistic case. We show how to give these kinematical algebras an infinite-dimensional lift by imposing suitable boundary conditions in a theory we call ``Carroll Gravity'', whose asymptotic symmetry algebra turns out to be an infinite-dimensional extension of the Carroll algebra.
}

\begin{document}
\maketitle
\flushbottom

\section{Introduction}
\label{sec:introduction}

Due to the principle of relativity, the notion of kinematical or space-time symmetry algebras, which contain all symmetries that relate different inertial frames, is a crucial ingredient in the construction of physical theories. Bacry and L\'evy-Leblond have classified all possibilities for kinematical algebras \cite{Bacry:1968zf}, consisting of space-time translations, spatial rotations and boosts, under some reasonable assumptions. Apart from the relativistic Poincar\'e and (A)dS algebras, this classification also contains the Galilei and Carroll algebras (and generalizations thereof that include a cosmological constant), that appear as kinematical algebras in the non-relativistic ($c \rightarrow \infty$) and ultra-relativistic ($c\rightarrow 0$) limit. Even though fundamental theories are relativistic, the Galilei and Carroll algebras continue to play an important role in current explorations of string theory and holography.

For instance, non-relativistic symmetries  underlie Newton-Cartan geometry, a differential geometric framework for non-relativistic space-times that has found recent applications in holography \cite{Son:2008ye, Kachru:2008yh, Balasubramanian:2008dm, Christensen:2013lma, Christensen:2013rfa, Hartong:2014oma, Hartong:2014pma, Bergshoeff:2014uea, Hartong:2015wxa}, Ho\v{r}ava-Lifshitz gravity \cite{Horava:2009uw, Hartong:2015zia, Hartong:2016yrf} and in the construction of effective field theories for strongly interacting condensed matter systems \cite{Son:2005rv, Hoyos:2011ez, Son:2013rqa, Abanov:2014ula, Gromov:2014vla, Gromov:2015fda, Geracie:2014nka, Festuccia:2016awg}.

On the other hand, ultra-relativistic Carroll symmetries have recently been studied in relation to their connection \cite{Duval:2014uva} with the Bondi--Metzner--Sachs (BMS) algebra of asymptotic symmetries of flat space-time \cite{Bondi:1962px, Sachs:1962zza}. As such, Carroll symmetries play a role in attempts to construct holographic dualities in asymptotically flat space-times \cite{Barnich:2010eb, Barnich:2012aw, Bagchi:2009my, Bagchi:2010eg, Bagchi:2014iea, Hartong:2015xda, Bagchi:2015wna, Hartong:2015usd, Bagchi:2016bcd}, as symmetries of the $S$-matrix in gravitational scattering \cite{Strominger:2013jfa} and in the recent notion of `soft hair' on black hole horizons \cite{Hawking:2016msc, Hawking:2016sgy}.\footnote{%
Note, however, that the near horizon boundary conditions in \cite{Donnay:2015abr, Donnay:2016ejv} lead to a symmetry algebra similar to but different from BMS, while the near horizon boundary conditions in \cite{Afshar:2016wfy,Grumiller:2016kcp,Afshar:2016kjj} lead to infinite copies of the Heisenberg algebra, in terms of which BMS (or related symmetry algebras) are composite.
}

The kinematical algebras that have been classified by Bacry and L\'evy-Leblond pertain to theories that contain bosonic fields with spins up to 2. One can also consider theories in which massless higher spin fields are coupled to gravity \cite{Vasiliev:1990en}. These so-called `higher spin gauge theories' have been formulated in (A)dS space-times (see \cite{Bekaert:2005vh, Vasiliev:2012vf, Didenko:2014dwa} for reviews) and have featured prominently in the AdS/CFT literature, as a class of theories for which holographic dualities can be constructed rigorously \cite{Giombi:2009wh, Giombi:2010vg, Giombi:2012ms, Gaberdiel:2010pz, Gaberdiel:2011zw, Candu:2012jq, Gaberdiel:2012uj, Candu:2012ne, Beccaria:2013wqa}, essentially because they are a `weak-weak' type of duality, i.e., CFTs with unbroken higher spin currents are free \cite{Maldacena:2011jn}. They typically contain an infinite number of higher spin fields. As a consequence, their space-time symmetries are extended to infinite-dimensional algebras that include higher spin generalizations of space-time translations, spatial rotations and boosts. Higher spin gauge theories have thus far mostly been considered in relativistic (A)dS space-times, with relativistic CFT duals\footnote{See however \cite{Gary:2012ms, Afshar:2012nk, Gutperle:2013oxa, Gary:2014mca, Breunhoelder:2015waa, Lei:2015ika,Lei:2015gza} for attempts to consider higher spin theories in non-AdS backgrounds with non-relativistic CFT duals.}.

Since both higher spin gauge theories as well as non- and ultra-relativistic space-time symmetries have played an important role in recent developments in holography, it is natural to ask whether one can combine the two. In order to answer this question, one needs to know which non- and ultra-relativistic kinematical algebras can appear as space-time symmetries of higher spin theories. This is the problem that we will start addressing in this paper, in the context of higher spin gauge theories in three space-time dimensions.

The reason for restricting ourselves to three space-time dimensions stems from the fact that, as far as higher spin gauge theory is concerned, this case is a lot simpler than its higher-dimensional counterpart. For instance, in three dimensions it is  possible to consider higher spin gauge theory in flat space-times \cite{Afshar:2013vka, Gonzalez:2013oaa, Grumiller:2014lna, Gary:2014ppa, Matulich:2014hea}, unlike the situation in higher dimensions where a non-zero cosmological constant is required\footnote{See however \cite{Sleight:2016dba,Sleight:2016xqq,Ponomarev:2016lrm} for recent progress concerning higher spin theories in four dimensional flat space.}. Moreover, in three dimensions higher spin gauge theories with only a finite number of higher spin fields can be constructed \cite{Aragone:1983sz}. In the relativistic case, such theories assume the form of Chern--Simons theories, for a gauge group that is a suitable finite-dimensional extension of the three-dimensional (A)dS and Poincar\'e groups. For theories with integer spins ranging from 2 to $N$ in AdS space-time, this gauge group is given by $\mathrm{SL}(N,\R) \otimes \mathrm{SL}(N,\R)$. Here, we will restrict ourselves for simplicity to `spin-3 theory' for which $N=3$, although our analysis can straightforwardly be generalized to arbitrary $N$.

In this paper, we will thus extend the discussion of kinematical algebras of \cite{Bacry:1968zf} to theories in three space-time dimensions that include a spin-3 field coupled to gravity. In particular, we will start from the observation made in \cite{Bacry:1968zf} that all kinematical algebras can be obtained by taking sequential In\"on\"u-Wigner (IW) contraction procedures\footnote{The terminology `IW contraction procedures' might perhaps sound a little unconventional at this point. We refer to section \ref{sec:allkinalgs} for a more precise discussion about the difference in our use of the terms `contraction' and `contraction procedures' and why this is relevant for our work.} of the (A)dS algebras. We will then classify all possible IW contraction procedures of the kinematical algebra of spin-3 theory in (A)dS$_3$, as well as all possible kinematical algebras that can be obtained by sequential contraction procedures. Some of the kinematical algebras that are obtained in this way can be interpreted as spin-3 extensions of the Galilei and Carroll algebras. We will show that one can construct Chern--Simons theories for (suitable extensions of) these algebras. These can then be interpreted as non- and ultra-relativistic three-dimensional spin-3 theories. We will in particular argue that these theories can be viewed as higher spin generalizations of Extended Bargmann gravity \cite{Papageorgiou:2009zc,Papageorgiou:2010ud,Bergshoeff:2016lwr,Hartong:2016yrf} and Carroll gravity \cite{Bergshoeff:2017btm}, two examples of non- and ultra-relativistic gravity theories that have been considered in the literature recently.

The kinematical algebras of spin-3 theories that we obtain in this paper are finite-dimensional. Relativistic three-dimensional kinematical algebras have infinite-dimensional extensions that are obtained as asymptotic symmetry algebras upon imposing suitable boundary conditions on metric and higher spin fields, such as the Virasoro algebra (for the AdS algebra) \cite{Brown:1986nw}, the BMS algebra (for the Poincar\'e algebra) \cite{Ashtekar:1996cd, Barnich:2006av} or $W$-algebras (for their higher spin generalizations) \cite{Henneaux:2010xg, Campoleoni:2010zq}.
It is interesting to ask whether the non- and ultra-relativistic algebras found in this paper also have infinite-dimensional extensions that correspond to asymptotic symmetry algebras of their corresponding higher spin gravity theories. We will not attempt to address this question in full generality in this paper. We will, however, show that the spin-2 Carroll algebra allows for an infinite-dimensional extension. In particular, we will show that there exist suitable boundary conditions in three-dimensional Carroll gravity, such that the resulting asymptotic symmetry algebra is an infinite-dimensional extension of the Carroll algebra. This suggests that a similar result should also hold for the non- and ultra-relativistic spin-3 theories constructed in this paper as well as for the other spin-2 theories that have not been investigated in detail yet.

The organization of this paper is as follows. In section \ref{sec:algebras}, we classify all IW contraction procedures of the kinematical algebra of spin-3 theory in (A)dS$_3$. We then classify all kinematical algebras that can be obtained by combining these various contraction procedures. In section \ref{sec:spin-3}, we restrict ourselves to the algebras that can be interpreted as non- and ultra-relativistic ones, for zero cosmological constant. We argue that in the ultra-relativistic cases, a Chern--Simons theory can be constructed in a straightforward manner. This is not true for the non-relativistic cases. However, we demonstrate that the non-relativistic kinematical algebras can be suitably extended in such a way that a Chern--Simons action can be written down. We then show via a linearized analysis that the non- and ultra-relativistic spin-3 Chern--Simons theories thus obtained can be viewed as spin-3 generalizations of Extended Bargmann gravity and Carroll gravity, respectively. In section \ref{sec:4} we discuss boundary conditions for Carroll gravity that lead to an infinite-dimensional extension of the Carroll algebra. This section does not depend on the results of the previous sections and can therefore be read independently. Finally, in \ref{sec:discussion} we end with our conclusions and an outlook for future work.

\section{Kinematical Spin-3 Algebras}
\label{sec:algebras}

In this section, we will be concerned with three-dimensional kinematical spin-3 algebras, i.e.\ generalized space-time symmetry algebras of theories of interacting, massless spin-2 and spin-3 fields. In particular, following Bacry and L\'evy--Leblond~\cite{Bacry:1968zf} we will classify all such algebras that can be obtained by combining different In\"on\"u-Wigner (IW) contraction~\cite{Inonu:1953sp} procedures from the algebras that underlie spin-3 gravity in AdS$_3$ and dS$_3$. After recalling the latter, we will present all possible ways of contracting them, such that non-trivial kinematical spin-3 algebras are obtained, via a classification theorem. The proof of this theorem is relegated to appendix \ref{app:proof}. Combining different of these contraction procedures leads to various kinematical spin-3 algebras, some of which will be discussed in the next section as a starting point for considering Carroll and Galilei spin-3 gravity Chern--Simons theories.

\subsection{AdS$_3$ and dS$_3$ Spin-3 Algebras}
\label{ssec:AdSdS}

Spin-3 gravity in (A)dS$_3$ \cite{Henneaux:2010xg,Campoleoni:2010zq} can be written as a Chern--Simons theory for the Lie algebra $\mathfrak{sl}(3,\R) \oplus \mathfrak{sl}(3,\R)$ (where the $\oplus$ denotes the direct sum as Lie algebras) for AdS$_3$ or $\mathfrak{sl}(3,\mathbb{C})$ (viewed as a real Lie algebra) for dS$_3$. In the following we will often denote the higher spin algebra  $\mathfrak{sl}(3,\R) \oplus \mathfrak{sl}(3,\R)$, realizing Spin-3 gravity in AdS$_{3}$, by $\mathfrak{hs}_{3}\mathfrak{AdS}$. Similarly, we indicate  the higher spin algebra $\mathfrak{sl}(3,\mathbb{C})$, realizing Spin-3 gravity in dS$_{3}$, by $\mathfrak{hs}_{3}\mathfrak{dS}$. In both cases, the algebra consists of the generators of Lorentz transformations $\hat \Jt_A$ and translations $\hat \Pt_A$ along with `spin-3 rotations' $\hat \Jt_{AB}$ and `spin-3 translations' $\hat \Pt_{AB}$, that are traceless-symmetric in the $(AB)$ indices ($A=0,1,2$) \footnote{We refer to appendix \ref{app:conventions} for index and other conventions used in this and upcoming sections.}:
\begin{align} \label{eq:JPsymmtraceless}
   & \hat\Jt_{AB} =\hat\Jt_{BA}\,,     & \eta^{AB}\hat \Jt_{AB} = 0 \,, \nonumber
\\
   & \hat\Pt_{AB}=\hat\Pt_{BA} \,,     & \eta^{AB}\hat \Pt_{AB} = 0 \,.
\end{align}
Here, $\eta^{AB}$ is the three-dimensional Minkowski metric.
We will often refer to $\{\hat \Jt_A,\hat \Pt_A\}$ as the `spin-2 generators' or the `spin-2 part' and similarly to $\{\hat \Jt_{AB}, \hat \Pt_{AB}\}$ as the `spin-3 generators' or `spin-3 part'.
Their commutation relations are given by~\cite{Henneaux:2010xg,Campoleoni:2010zq}
\begin{align} \label{eq:AdSdSalgebra}
 \left[\,\hat \Jt_A \comma \hat \Jt_B \,\right]  &=  \epsilon_{ABC} \,\hat \Jt^C \,,  & \left[\,\hat \Jt_A \comma \hat \Pt_B \,\right]  &=  \epsilon_{ABC} \,\hat \Pt^C\,, \nonumber \\
 \left[\,\hat \Pt_A \comma \hat \Pt_B \,\right] & = \pm  \epsilon_{ABC} \,\hat \Jt^C \,,  & & \nonumber \\
 \left[\, \hat \Jt_A \comma \hat \Jt_{BC} \,\right] & =  \epsilon\indices{^M_{A(B}} \, \hat \Jt_{C)M} \,,
      & \left[\, \hat \Pt_A \comma \hat \Pt_{BC} \,\right]  &=  \pm 
        \, \epsilon\indices{^M_{A(B}} \, \hat \Jt_{C)M} \,, \nonumber
    \\
 \left[\, \hat \Jt_A \comma \hat \Pt_{BC} \,\right]  &=  \epsilon\indices{^M_{A(B}}\, \hat \Pt_{C)M} \,,
 &  \left[\, \hat \Pt_A \comma \hat \Jt_{BC} \,\right] & =  \epsilon\indices{^M_{A(B}} \, \hat \Pt_{C)M} \,, \nonumber \\
\left[\, \hat \Jt_{AB} \comma\hat \Jt_{CD} \,\right]  &=  -  \eta_{(A(C} \epsilon_{D)B)M}  \, \hat \Jt^M \,,
   & \left[\,\hat \Jt_{AB} \comma \hat \Pt_{CD} \,\right]  &=  -  \eta_{(A(C} \epsilon_{D)B)M} \, \hat \Pt^M \,, \nonumber
    \\
    \left[\, \hat \Pt_{AB} \comma \hat \Pt_{CD} \,\right] & =  \mp  
      \eta_{(A(C} \epsilon_{D)B)M} \, \hat \Jt^M  \,,
\end{align}
where the upper sign refers to $\mathfrak{hs}_{3}\mathfrak{AdS}$ and the lower sign to $\mathfrak{hs}_{3}\mathfrak{dS}$. Note that the first two lines constitute the isometry algebra of (A)dS$_3$, i.e.\ $\mathfrak{sl}(2,\R) \oplus \mathfrak{sl}(2,\R)$ for AdS$_3$ and $\mathfrak{sl}(2,\mathbb{C})$, viewed as a real Lie algebra, for dS$_3$.

For future reference, we mention that the above algebras are equipped with a nondegenerate ad-invariant symmetric bilinear form that we will denote by $\langle \cdot\,, \cdot\rangle$ (where the $\cdot$ are placeholders for generators) and that we will henceforth call the `invariant metric'. Its non-zero components are given by
\begin{equation} \label{eq:bilformAdSdS}
  \langle \hP_A\,, \hJ_B\rangle = \eta_{AB} \,, \qquad \langle \hP_{AB}\,, \hJ_{CD}\rangle = \eta_{A(C} \eta_{D)B} - \frac23 \eta_{AB} \eta_{CD} \,.
\end{equation}
Note that this represents an invariant metric for both $\mathfrak{hs}_{3}\mathfrak{AdS}$ and $\mathfrak{hs}_{3}\mathfrak{dS}$. The existence of this metric allows one to construct Chern--Simons actions for the algebras $\mathfrak{hs}_{3}\mathfrak{AdS}$ and $\mathfrak{hs}_{3}\mathfrak{dS}$, that correspond to the actions for spin-3 gravity in (A)dS$_3$ \cite{Henneaux:2010xg,Campoleoni:2010zq}.

In the following, it will prove convenient to introduce a time-space splitting of the indices $A = \{0,a; a=1,2\}$. We will thereby   use the following notation:
\begin{align}
  \label{eq:nothsalgebras}
\Jt &= \hat \Jt_0 \,, & \Gt_a &= \hat \Jt_a \,, & \Ht &= \hat \Pt_0 \,, & \Pt_a &= \hat \Pt_a\,, \nonumber \\
\Jt_a &= \hat \Jt_{0a} \,, & \Gt_{ab} &= \hat \Jt_{ab} \,, & \Ht_a &= \hat \Pt_{0a} \,, & \Pt_{ab} &= \hat \Pt_{ab}\,.
\end{align}
Note that we have left out the generators $\hat{\Pt}_{00}$ and $\hat{\Gt}_{00}$ here. These generators are not independent, due to the tracelessness constraint (\ref{eq:JPsymmtraceless}) and in the following we will eliminate them in favour of $\Pt_{ab}$ and $\Gt_{ab}$. After these substitutions, the commutation relations of $\mathfrak{hs}_{3}\mathfrak{(A)dS}$ in this new basis are given in the first column of table~\ref{tab:adspoin}.

\subsection{All Kinematical Spin-3 Algebras by Contracting $\mathfrak{hs}_{3}\mathfrak{(A)dS}$}
\label{sec:allkinalgs}

Before discussing spin-3, it is convenient to start with giving  a short review of the spin-2 case~\cite{Bacry:1968zf}.
Since both the spin-2 and spin-3 cases make use of the IW contraction we first discuss this procedure.
We will use this as an opportunity to introduce some terminology that will be used throughout this paper.

Starting from a Lie algebra $\mathfrak{g}$, one can choose a subalgebra $\mathfrak{h}$ and consider the decomposition $\mathfrak{g} = \mathfrak{h} + \mathfrak{i}$
where $+$ denotes the direct sum as vector spaces (not as Lie algebras). Upon rescaling the generators of $\mathfrak{i}$ with a so-called contraction parameter $\epsilon$, $\mathfrak{i} \rightarrow \epsilon \mathfrak{i}$, the commutation relations of $\mathfrak{g}$ assume the following form
\begin{align}
  [\,\mathfrak{h} \comma \mathfrak{h} \,] & \subseteq \mathfrak{h} \,, &
  [\,\mathfrak{h} \comma \mathfrak{i}\,] & \subseteq \frac{1}{\epsilon}\, \mathfrak{h} + \mathfrak{i} \,, &
  [\,\mathfrak{i} \comma \mathfrak{i}\,] & \subseteq \frac{1}{\epsilon^{2}}\,\mathfrak{h} +\frac{1}{\epsilon}\, \mathfrak{i} \,.
\end{align}
One thus sees that the limit $\epsilon \rightarrow \infty$ is well-defined\footnote{For this to be true, it is crucial that $\mathfrak{h}$ is chosen as a subalgebra of $\mathfrak{g}$. Indeed, for $\mathfrak{h}$ a generic subspace of $\mathfrak{g}$, one has upon rescaling the generators of $\mathfrak{i}$ that $[\,\mathfrak{h} \comma \mathfrak{h} \,] \subseteq \mathfrak{h} + \epsilon\, \mathfrak{i}$ and the limit $\epsilon \rightarrow \infty$ is no longer well-defined.}. Taking this limit leads to an inequivalent algebra, that is a semidirect sum $\mathfrak{i} \niplus \mathfrak{h}$, for which $\mathfrak{i}$ is an abelian ideal
\begin{align} \label{eq:IWstructure}
  [\,\mathfrak{h} \comma \mathfrak{h} \,] & \subseteq \mathfrak{h} \,, &
  [\,\mathfrak{h} \comma \mathfrak{i}\,] & \subseteq  \mathfrak{i} \,, &
  [\,\mathfrak{i} \comma \mathfrak{i}\,] & = 0 \,.
\end{align}
This algebra is called the IW contraction of $\mathfrak{g}$ with respect to $\mathfrak{h}$. The IW contraction is called `trivial' if either $\mathfrak{h} = \mathfrak{g}$ or $\mathfrak{i} = \mathfrak{g}$. The procedure that leads to an IW contraction, i.e.\ that consists of choosing a subalgebra $\mathfrak{h}$, rescaling the generators of $\mathfrak{i}$ as $\mathfrak{i} \rightarrow \epsilon \, \mathfrak{i}$ and taking $\epsilon \rightarrow \infty$, will be denoted as the `IW contraction procedure' in this paper.

Note that a nontrivial IW contraction procedure is uniquely specified by a suitable choice of the subalgebra $\mathfrak{h} \subset \mathfrak{g}$. Not all possible subalgebras, however, lead to interesting IW contractions that can e.g.\ be interpreted as kinematical algebras. For spin-2, the question which contraction procedures of the isometry algebras of AdS or dS lead to kinematical algebras, has been addressed by Bacry and L\'evy--Leblond~\cite{Bacry:1968zf}. In particular, they have shown that there are only four different IW contraction procedures of the AdS or dS algebras that lead to kinematical algebras. These have been called `space-time', `speed-space', `speed-time' and `general' in \cite{Bacry:1968zf}.
Effectively, the first three of these contractions can be described by either taking a limit of the (A)dS radius $\ell$ or the speed of light $c$. Specifically, the space-time contraction corresponds to $\ell \to \infty$, the speed-time contraction corresponds to $c \rightarrow 0$ and the speed-space contraction corresponds to $c\rightarrow \infty$. However, in this work we suppress factors of $\ell$ and $c$.
The general contraction procedure can also be obtained as a sequential contraction of the other three and therefore does not provide us with a new algebra.
Moreover, it has been shown that there are in total 8 possible kinematical algebras\footnote{The possible kinematical algebras considered in \cite{Bacry:1968zf} are all possible space-time symmetry algebras that obey the assumptions that space is isotropic and therefore their generators have the correct ($\Ht$ is a scalar, $\Pt,\Jt,\Gt$ are vectors) transformation behavior under rotations. Furthermore, parity and time-reversal are automorphisms and  boosts are non-compact.} that can be obtained by combining different IW contraction procedures of the AdS or dS isometry algebras. We have summarized the four IW contraction procedures in the following table \ref{tab:spin2contr}, by indicating the subalgebra $\mathfrak{h}$ with respect to which the contraction procedure is taken, as well as the generators that form the abelian ideal $\mathfrak{i}$:
\begin{table}[H]
  \centering
$
  \begin{array}{l l l l l }
\toprule
    \text{Contraction } & \phantom{aa} & \multicolumn{1}{c}{\mathfrak{h}} & \phantom{aa} & \multicolumn{1}{c}{\mathfrak{i}} \\ \midrule
    \text{Space-time}   &              & \{\Jt,\Gt_{a} \}                 &              & \{\Ht,\Pt_{a} \}                 \\
    \text{Speed-space}  &              & \{\Jt,\Ht \}                     &              & \{\Gt_{a},\Pt_{a} \}             \\
    \text{Speed-time}   &              & \{\Jt, \Pt_{a} \}                &              & \{\Gt_{a}, \Ht \}                \\
     \text{General}     &              & \{\Jt \}                         &              & \{\Ht, \Pt_{a}, \Gt_{a} \}       \\ \bottomrule
  \end{array}
$
\caption{The four different IW contraction procedures classified in~\cite{Bacry:1968zf}.}
  \label{tab:spin2contr}
\end{table}
\noindent The names of the eight kinematical algebras of \cite{Bacry:1968zf}, along with the symbols we will use to denote them, are given in table \ref{tab:kinsp2}.
\begin{table}[h]
  \centering
  \begin{tabular}{l  l} \toprule
    Name & Symbol\\ \midrule
    (Anti) de Sitter & $\mathfrak{(A)dS}$\\
    Poincar\'e & $\mathfrak{poi}$\\
    Para-Poincar\'e & $\mathfrak{ppoi}$\\
    Newton--Hooke & $\mathfrak{nh}$\\
    Galilei & $\mathfrak{gal}$\\
    Para-Galilei & $\mathfrak{pgal}$\\
    Carroll & $\mathfrak{car}$\\
    Static & $\mathfrak{st}$\\\bottomrule
  \end{tabular}
  \caption{Names of the kinematical algebras and the symbols that denote them.}
  \label{tab:kinsp2}
\end{table}

\begin{figure}[h]
  \centering
\tdplotsetmaincoords{80}{120}
\begin{tikzpicture}[
tdplot_main_coords,
dot/.style={circle,fill},
linf/.style={ultra thick,->,blue},
cinf/.style={ultra thick,->,red},
tinf/.style={ultra thick,->},
stinf/.style={ultra thick,->,gray},
scale=0.7
]

\node (ads) at (0,10,10) [dot, label=above:$\mathfrak{(A)dS}$] {};
\node (p) at (10,10,10) [dot, label=above:$\mathfrak{poi}$] {};
\node (nh) at (0,10,0) [dot, label=below:$\mathfrak{nh}$] {};
\node (pp) at (0,0,10) [dot, label=above:$\mathfrak{ppoi}$] {};
\node (g) at (10,10,0) [dot, label=below:$\mathfrak{gal}$] {};
\node (pg) at (0,0,0) [dot, label=below:$\mathfrak{pgal}$] {};
\node (car) at (10,0,10) [dot, label=above:$\mathfrak{car}$] {};
\node (st) at (10,0,0) [dot, label=below:$\mathfrak{st}$] {};

\draw[linf] (ads) -- node [sloped,above,xshift=-4mm] {Space-time
}  (p);

\draw[linf] (nh) -- (g);
\draw[linf] (pp) -- (car);
\draw[linf,dashed] (pg) -- (st);

\draw[cinf,double] (ads) -- node [sloped,above] {Speed-space 
} (nh);
\draw[cinf,double,dashed] (pp) -- (pg);
\draw[cinf] (p) -- (g);
\draw[cinf] (car) -- (st);

\draw[tinf,double] (ads) -- node [sloped,above] {Speed-time
} (pp);
\draw[tinf] (p) -- (car);
\draw[tinf] (g) -- (st);
\draw[tinf,double,dashed] (nh) -- (pg);

\draw[stinf,dashed] (ads) -- node [sloped,above] {General}(st);
\end{tikzpicture}
  \caption{This cube summarizes the sequential contractions starting from $\mathfrak{(A)dS}$. The lines represent contraction procedures and the dots represent the resulting contractions. We consider contraction procedures starting from AdS and dS simultaneously. Each dot can therefore represent one contraction, if the contraction procedures from AdS and dS lead to the same algebra, or two contractions, if the contraction procedures from AdS and dS lead to two different results. We have indicated this in the cube by using single lines, for contraction procedures that lead to the same contraction, and double lines otherwise. Dashed lines have no specific meaning except that they should convey the feeling of a three-dimensional cube.}
  \label{fig:cube}
\end{figure}
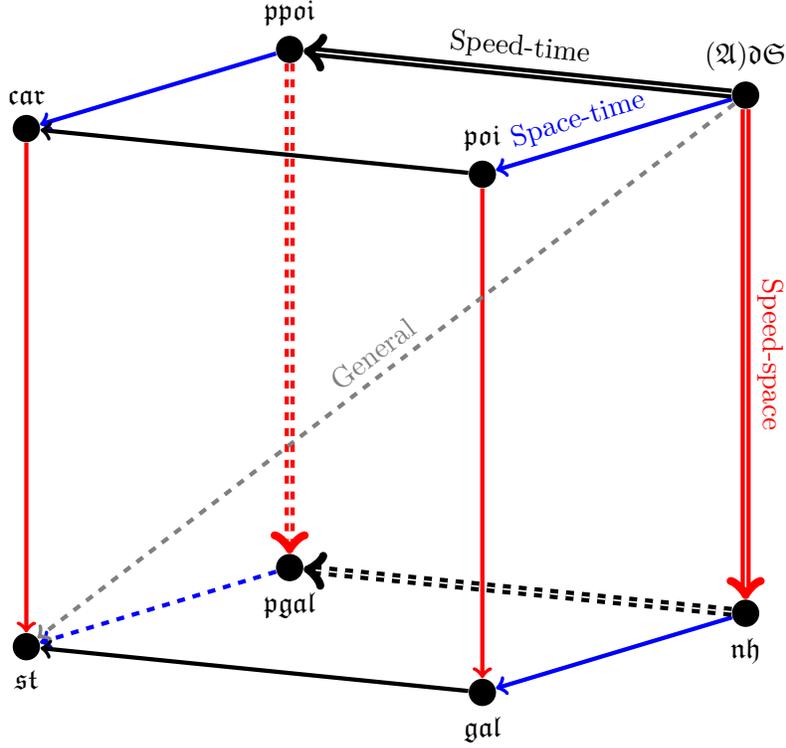

\noindent The  IW contraction procedures and the contracted algebras that we discussed so far can be conveniently summarized as a cube, see figure \ref{fig:cube}.

We next consider the spin-3 case where, following the spin-2 case, we will obtain a classification of all possible contraction procedures\footnote{Here, we will classify different contraction procedures, in the sense defined above as different choices of subalgebra $\mathfrak{h}$. This does not mean that all these contraction procedures lead to non-isomorphic Lie algebras. Indeed, in the analysis of \cite{Bacry:1968zf} e.g. one can see that the space-time and speed-time contraction procedures applied to the AdS$_3$ isometry algebra lead to two contractions that are both isomorphic to the Poincar\'e algebra. We should however mention that these algebras are isomorphic in the mathematical sense; physically they can be regarded as non-equivalent as the isomorphism that relates them corresponds to an interchange of boost and translation generators. Note also that the different contraction procedures that are classified here are not necessarily independent. As an example, one can check that the general contraction procedure of table \ref{tab:spin2contr} can be obtained by sequential space-time, speed-space and speed-time contractions in an arbitrary order.} of $\mathfrak{hs}_{3}\mathfrak{AdS}$ and $\mathfrak{hs}_{3}\mathfrak{dS}$ by listing all their possible subalgebras. We start from $\mathfrak{hs}_3\mathfrak{(A)dS}$ since these are semisimple algebras and can therefore not be viewed as contractions themselves (as nontrivial contraction procedures always lead to algebras with an abelian ideal that are thus not semisimple). Now, in order to obtain contractions that can be identified as interesting kinematical spin-3 algebras, we will impose two restrictions:
\begin{itemize}
\item When restricted to the spin-2 part of the algebra, the contraction procedures should correspond to those considered in table \ref{tab:spin2contr}. This ensures that the spin-2 parts of the algebras obtained by various combinations of these contraction procedures correspond to the kinematical algebras of \cite{Bacry:1968zf}.
\item Furthermore, we will also demand that in the resulting contraction not all commutators of the spin-3 part are vanishing. This requirement is motivated by the fact that we are interested in using these contractions to describe fully interacting theories of massless spin-2 and spin-3 fields. Indeed, as we will show later on, for some of the algebras obtained here, one can construct a Chern--Simons action for spin-2 and spin-3 fields. Only when the commutators of the spin-3 part are not all vanishing, do the spin-3 fields contribute to the equations of motion of the spin-2 fields.
\end{itemize}
All ways of contracting $\mathfrak{hs}_{3}\mathfrak{AdS}$ and $\mathfrak{hs}_{3}\mathfrak{dS}$ that obey these two restrictions can then be summarized by the following theorem:
\begin{theorem}
All possible IW contraction procedures, that reduce to those considered in table \ref{tab:spin2contr} when restricted to the spin-2 part and that are non-abelian on the subspace spanned by the spin-3 generators $\{\Jt_{a},\Ht_{a},\Gt_{ab},\Pt_{ab} \}$, are given by 10 `democratic' contraction procedures that are specified in table \ref{tab:contr} and 7 `traceless' contraction procedures, given in table \ref{tab:excontr}. As in table \ref{tab:spin2contr}, we have specified these contraction procedures by indicating the subalgebra $\mathfrak{h}$ with respect to which $\mathfrak{hs}_{3}\mathfrak{(A)dS}$ is contracted, as well as by giving the resulting abelian ideal $\mathfrak{i}$.

\begin{table}[H]
  \centering
$
  \begin{array}{l r l l }
\toprule
    \text{Contraction }                                                                                                & \# & \multicolumn{1}{c}{\mathfrak{h}}    & \multicolumn{1}{c}{\mathfrak{i}}                     \\ \midrule
    \text{Space-time}                                                                                                  & 1  & 
                                           \{\Jt, \Gt_{a},\Jt_{a},\Gt_{ab}\}                                           & 
                                                                                \{\Ht, \Pt_{a},\Ht_{a}, \Pt_{ab} \}                                                                                                      \\
                                                                                                                       & 2  & 
                                                                                    \{\Jt, \Gt_{a},\Ht_{a}, \Pt_{ab}\} & 
                                                                                                                         \{\Ht, \Pt_{a},\Jt_{a},\Gt_{ab} \}                                                              \\ \midrule
    \text{Speed-space}                                                                                                 & 3  & \{\Jt,\Ht,\Jt_{a}, \Ht_{a} \}       & \{\Gt_{a},\Pt_{a},\Gt_{ab},\Pt_{ab} \}               \\
                                                                                                                       & 4  & \{\Jt,\Ht,\Gt_{ab},\Pt_{ab} \}      & \{\Gt_{a},\Pt_{a},\Jt_{a}, \Ht_{a} \}                \\ \midrule
    \text{Speed-time}                                                                                                  & 5  & \{\Jt, \Pt_{a},\Jt_{a},\Pt_{ab} \}  & \{\Gt_{a}, \Ht,\Ht_{a},\Gt_{ab} \}                   \\
                                                                                                                       & 6  & \{\Jt, \Pt_{a},\Ht_{a},\Gt_{ab}  \} & \{\Gt_{a}, \Ht,\Jt_{a},\Pt_{ab} \}                   \\ \midrule
                                                                                                                       & 7  & \{\Jt,\Jt_{a} \}                    & \{\Ht, \Pt_{a}, \Gt_{a},\Ht_{a},\Gt_{ab},\Pt_{ab} \} \\
     \text{General}                                                                                                    & 8  & \{\Jt,\Gt_{ab} \}                   & \{\Ht, \Pt_{a}, \Gt_{a},\Jt_{a},\Ht_{a} ,\Pt_{ab} \} \\
                                                                                                                       & 9  & \{\Jt,\Ht_{a} \}                    & \{\Ht, \Pt_{a}, \Gt_{a},\Jt_{a},\Gt_{ab},\Pt_{ab} \} \\
                                                                                                                       & 10 & \{\Jt,\Pt_{ab} \}                   & \{\Ht, \Pt_{a}, \Gt_{a},\Jt_{a},\Ht_{a} ,\Gt_{ab} \} \\ \bottomrule
  \end{array}
$
  \caption{All democratic contraction procedures}
  \label{tab:contr}
\end{table}

\begin{table}[H]
  \centering
$
  \begin{array}{l r r l l }
\toprule
    \text{Contraction} & \#  & \multicolumn{1}{c}{\mathfrak{h}}                                     & \multicolumn{1}{c}{\mathfrak{i}}                                                          \\ \midrule
    \text{Speed-space} & 4a  & \{\Jt,\Ht,\Gt_{ab},\Pt_{12}, \Pt_{22}-\Pt_{11} \}                    & \{\Gt_{a},\Pt_{a},\Jt_{a}, \Ht_{a}, \Pt_{11}+\Pt_{22} \}                                  \\
                       & 4b  & \{\Jt,\Ht,\Gt_{12}, \Gt_{22}-\Gt_{11}, \Pt_{ab} \}                   & \{\Gt_{a},\Pt_{a},\Jt_{a}, \Ht_{a},\Gt_{11}+\Gt_{22} \}                                   \\
                       & 4c  & \{\Jt,\Ht,\Gt_{12}, \Gt_{22}-\Gt_{11},\Pt_{12}, \Pt_{22}-\Pt_{11} \} & \{\Gt_{a},\Pt_{a},\Jt_{a}, \Ht_{a},\Gt_{11}+\Gt_{22}, \Pt_{11}+\Pt_{22} \}                \\ \midrule
                       & 8a  & \{\Jt,\Gt_{12}, \Gt_{22}-\Gt_{11} \}                                 & \{\Ht, \Pt_{a}, \Gt_{a},\Jt_{a},\Ht_{a}, \Gt_{11}+\Gt_{22},\Pt_{ab} \}                    \\
     \text{General}    & 10a & \{\Jt,\Pt_{12}, \Pt_{22}-\Pt_{11} \}                                 & \{\Ht, \Pt_{a}, \Gt_{a},\Jt_{a},\Ht_{a} ,\Gt_{ab},\Pt_{11}+\Pt_{22} \}                    \\
                       & 8b  & \{\Jt,\Gt_{12}, \Gt_{22}-\Gt_{11},\Pt_{11}+\Pt_{22} \}               & \{\Ht, \Pt_{a}, \Gt_{a},\Jt_{a},\Ht_{a}, \Gt_{11}+\Gt_{22},\Pt_{12}, \Pt_{22}-\Pt_{11} \} \\
                       & 10b & \{\Jt,\Pt_{12}, \Pt_{22}-\Pt_{11}, \Gt_{11}+\Gt_{22}\}               & \{\Ht, \Pt_{a}, \Gt_{a},\Jt_{a},\Ht_{a} ,\Gt_{12}, \Gt_{22}-\Gt_{11},\Pt_{11}+\Pt_{22} \} \\ \bottomrule
  \end{array}
$
  \caption{All traceless contraction procedures}
  \label{tab:excontr}
\end{table}
\end{theorem}

A complete proof of this theorem is given in appendix \ref{app:proof}. For now, let us suffice by saying that the proof starts by noting that each of the subalgebras $\mathfrak{h}$ in table \ref{tab:spin2contr} needs to be supplemented with spin-3 generators, in order to have a contraction with a non-abelian spin-3 part. The proof then proceeds by enumerating, for each of the contraction procedures of table \ref{tab:spin2contr}, all possibilities in which spin-3 generators can be added to $\mathfrak{h}$ such that one still obtains a subalgebra, that leads to a contraction with a non-abelian spin-3 part. We refer to appendix \ref{app:all-expl-contr} for the explicit Lie algebras of the contraction procedures given in table \ref{tab:contr}.

Finally, let us  comment on the terminology `democratic' and `traceless'. This terminology stems from the fact that the three independent generators contained in $\Pt_{ab}$ ($\Gt_{ab}$) form a real, reducible representation of $\Jt$, that can be split into a tracefree symmetric part consisting of the generators $\{\Pt_{12}, \Pt_{22}-\Pt_{11}\}$ ($\{\Gt_{12}, \Gt_{22}-\Gt_{11}\}$) and a trace part $\Pt_{11} + \Pt_{11}$ ($\Gt_{11} + \Gt_{22}$). The democratic contraction procedures are such that the subalgebra  $\mathfrak{h}$ contains both tracefree symmetric and trace components of $\Pt_{ab}$ ($\Gt_{ab}$), if present. In some cases, it is not necessary to include the trace component in $\mathfrak{h}$ in order to obtain a valid subalgebra. This is the case for the democratic contraction procedures, numbered 4, 8 and 10 in table \ref{tab:contr}. Moving the trace component from $\mathfrak{h}$ to $\mathfrak{i}$ leads to the traceless cases $4a$, $4b$, $4c$, $8a$ and $10a$ in table \ref{tab:excontr}. In the last two remaining cases both the tracefree symmetric part of $\Gt_{ab}$ ($\Pt_{ab}$) and the trace part of $\Pt_{ab}$ ($\Gt_{ab}$) belong to the subalgebra  $\mathfrak{h}$.
Doing this leads to the traceless cases $8b$ and $10b$.

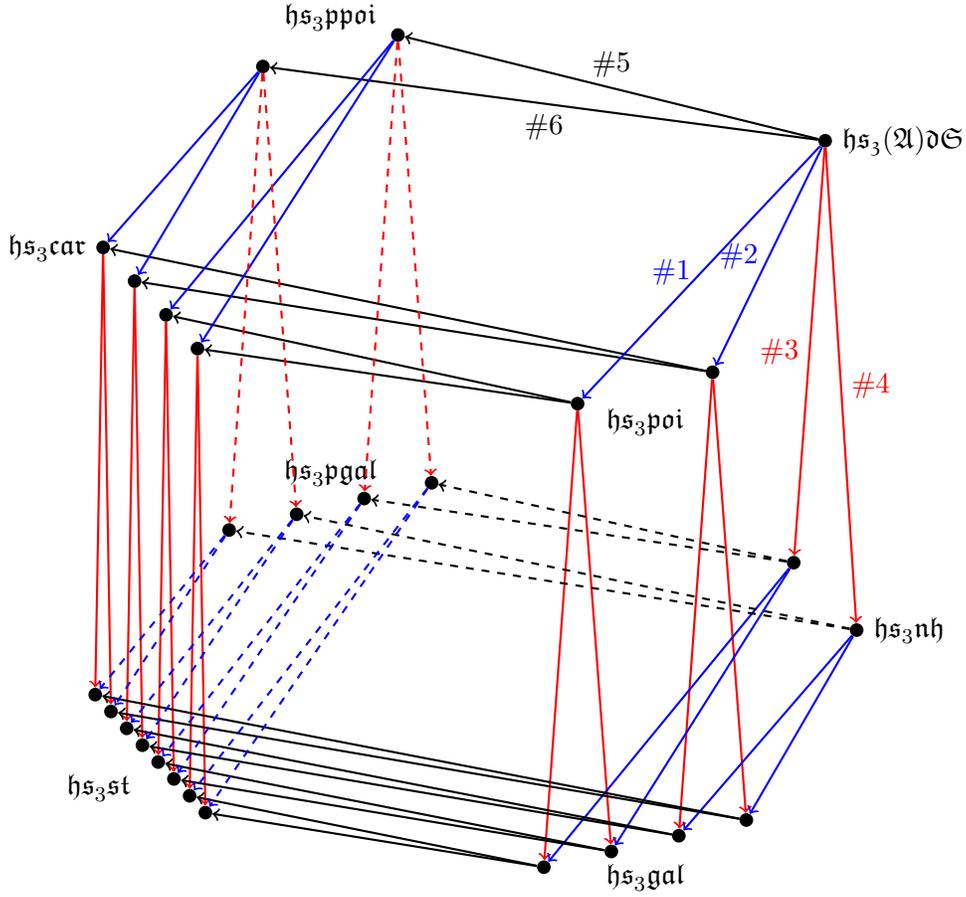
\begin{figure}[h]
  \centering
\tdplotsetmaincoords{60}{110}
\begin{tikzpicture}[
tdplot_main_coords,
dot/.style={circle,fill,scale=0.5},
linf/.style={thick,->,blue},
cinf/.style={thick,->,red},
tinf/.style={thick,->},
stinf/.style={ultra thick,->,gray},
scale=0.7
]


\node (ads) at (0,10,10) [dot, label=right:$\mathfrak{hs_3(A)}\mathfrak{dS}$] {};

\node (p) at (10,10,10) [label=below:$\mathfrak{hs}_{3}\mathfrak{poi}$] {};

\node (p1) at (10+\shi,10-\shi,10) [dot] {};
\node (p2) at (10-\shi,10+\shi,10) [dot] {};


\node (nh1) at (-\shi,10-\shi,0) [dot] {};
\node (nh2) at (\shi,10+\shi,0) [dot, label=right:$\mathfrak{hs}_{3}\mathfrak{nh}$] {};

\node (pp) at (0,0,10) [label=above:$\mathfrak{hs}_{3}\mathfrak{ppoi}$] {};

\node (pp1) at (-\shi,\shi,10) [dot] {};
\node (pp2) at (\shi,-\shi,10) [dot] {};

\node (g) at (10,10,0) [label=below:$\mathfrak{hs}_{3}\mathfrak{gal}$] {};

\node (g3) at (10-\shi-\shih,10+\shi+\shih,0) [dot] {};
\node (g4) at (10-\shi+\shih,10+\shi-\shih,0) [dot] {};

\node (g1) at (10+\shi+\shih,10-\shi-\shih,0) [dot] {};
\node (g2) at (10+\shi-\shih,10-\shi+\shih,0) [dot] {};

\node (pg) at (0,0,0) [label=above:$\mathfrak{hs}_{3}\mathfrak{pgal}$] {};


\node (pg1) at (-\shi+\shih,\shi-\shih,0) [dot] {};
\node (pg2) at (-\shi-\shih,\shi+\shih,0) [dot] {};


\node (pg3) at (\shi+\shih,-\shi-\shih,0) [dot] {};
\node (pg4) at (\shi-\shih,-\shi+\shih,0) [dot] {};


\node (car1) at (10+\shi-\shih,\shi-\shih,10) [dot] {};
\node (car2) at (10+\shi+\shih,\shi+\shih,10) [dot] {};

\node (car3) at (10-\shi-\shih,-\shi-\shih,10) [dot, label=left:$\mathfrak{hs}_{3}\mathfrak{car}$] {};
\node (car4) at (10-\shi+\shih,-\shi+\shih,10) [dot] {};

\node (st) at (10,0,0) [label=below left:$\mathfrak{hs}_{3}\mathfrak{st}$] {};

\node (st3) at (10-\shi-\shih-\shihh,-\shi-\shih-\shihh,0) [dot] {};
\node (st4) at (10-\shi-\shih+\shihh,-\shi-\shih+\shihh,0) [dot] {};
\node (st7) at (10+\shi+\shih+\shihh,\shi+\shih+\shihh,0) [dot] {};
\node (st8) at (10+\shi+\shih-\shihh,\shi+\shih-\shihh,0) [dot] {};

\node (st5) at (10+\shi-\shih+\shihh,\shi-\shih+\shihh,0) [dot] {};
\node (st6) at (10+\shi-\shih-\shihh,\shi-\shih-\shihh,0) [dot] {};

\node (st1) at (10-\shi+\shih+\shihh,-\shi+\shih+\shihh,0) [dot] {};
\node (st2) at (10-\shi+\shih-\shihh,-\shi+\shih-\shihh,0) [dot] {};

\draw[tinf] (p1) -- (car2);
\draw[tinf] (p1) -- (car1);
\draw[tinf] (p2) -- (car3);
\draw[tinf] (p2) -- (car4);

\draw[cinf] (ads) -- node [left] {\#3} (nh1);
\draw[cinf] (ads) -- node [right] {\#4}(nh2);

\draw[tinf] (ads) -- node [above] {\#5}(pp1);
\draw[tinf] (ads) -- node [below] {\#6}(pp2);

\draw[cinf] (p1) -- (g1);
\draw[cinf] (p1) -- (g2);
\draw[cinf] (p2) -- (g3);
\draw[cinf] (p2) -- (g4);

\draw[linf] (ads) -- node [left] {\#1} (p1);
\draw[linf] (ads) -- node [left] {\#2} (p2);

\draw[linf] (nh1) -- (g1);
\draw[linf] (nh1) -- (g2);
\draw[linf] (nh2) -- (g3);
\draw[linf] (nh2) -- (g4);

\draw[linf] (pp1) -- (car1);
\draw[linf] (pp1) -- (car2);
\draw[linf] (pp2) -- (car3);
\draw[linf] (pp2) -- (car4);

\draw[cinf] (car2) -- (st8);
\draw[cinf] (car2) -- (st7);

\draw[cinf] (car1) -- (st6);
\draw[cinf] (car1) -- (st5);

\draw[cinf] (car3) -- (st4);
\draw[cinf] (car3) -- (st3);
\draw[cinf] (car4) -- (st2);
\draw[cinf] (car4) -- (st1);

\draw[tinf,dashed] (nh1) -- (pg1);
\draw[tinf,dashed] (nh1) -- (pg2);
\draw[tinf,dashed] (nh2) -- (pg3);
\draw[tinf,dashed] (nh2) -- (pg4);

\draw[cinf,dashed] (pp1) -- (pg1);
\draw[cinf,dashed] (pp1) -- (pg2);
\draw[cinf,dashed] (pp2) -- (pg3);
\draw[cinf,dashed] (pp2) -- (pg4);

\draw[tinf] (g1) -- (st8);
\draw[tinf] (g1) -- (st7);
\draw[tinf] (g2) -- (st6);
\draw[tinf] (g2) -- (st5);
\draw[tinf] (g3) -- (st4);
\draw[tinf] (g3) -- (st3);
\draw[tinf] (g4) -- (st2);
\draw[tinf] (g4) -- (st1);

\draw[linf,dashed] (pg1) -- (st6);
\draw[linf,dashed] (pg1) -- (st5);
\draw[linf,dashed] (pg2) -- (st8);
\draw[linf,dashed] (pg2) -- (st7);

\draw[linf,dashed] (pg3) -- (st3);
\draw[linf,dashed] (pg3) -- (st4);
\draw[linf,dashed] (pg4) -- (st1);
\draw[linf,dashed] (pg4) -- (st2);

\end{tikzpicture}
  \caption{This figure summarizes the sequential democratic contractions of table \ref{tab:contr}. There are 2 space-time (blue; \#1,\#2), 2 speed-space (red; \#3,\#4) and 2 speed-time (black; \#5,\#6) contractions and combining them leads to the full cube. The commutators of the algebras corresponding to the dots are given in table \ref{tab:adspoin}-\ref{tab:hsstat2}. In comparison to figure \ref{fig:cube}, we have for clarity omitted the double lines and the diagonal lines that indicate the direct IW contraction procedures to the static algebras.}
  \label{fig:hscube}
\end{figure}

The democratic contractions can  again be summarized as a cube, see figure \ref{fig:hscube}.

\section{Spin-3 Carroll, Galilei and Extended Bargmann Chern--Simons Theories}
\label{sec:spin-3}

In the previous section, we have classified all possible contraction procedures of the spin-3 AdS$_3$ and dS$_3$ algebras. Combining some of these contraction procedures can lead to algebras whose spin-2 part corresponds to the Carroll or Galilei algebra. Here, we will study these cases in more detail. In particular, we will be concerned with constructing Chern--Simons theories for these spin-3 algebras, or suitable extensions thereof.
This extends \cite{Bergshoeff:2017btm} where the case of spin-2 Carroll and spin-2 Galilei gravity is discussed. 

Assuming that there is an invariant metric $\langle \cdot\,,\cdot \rangle$ available, the Chern--Simons action can be constructed as
\begin{equation}
  \label{eq:CSaction}
  S[A] = \int \langle A \wedge d A +\frac23 A \wedge A \wedge A \rangle \,,
\end{equation}
where $A$ is an algebra-valued gauge field. Using the invariant metric (\ref{eq:bilformAdSdS}), one can find the Chern--Simons actions for spin-3 gravity in (A)dS$_3$ \cite{Henneaux:2010xg,Campoleoni:2010zq}. Performing the first contraction procedure of table \ref{tab:contr}, i.e.\ rescaling $\hat{\Pt}_A \rightarrow \epsilon \, \hat{\Pt}_A$, $\hat{\Pt}_{AB} \rightarrow \epsilon \, \hat{\Pt}_{AB}$ and taking the limit $\epsilon \rightarrow \infty$, leads to the spin-3 Poincar\'e algebra $\mathfrak{hs}_{3}\mathfrak{poi1}$ of table \ref{tab:adspoin}. This algebra can be equipped with the same invariant metric (\ref{eq:bilformAdSdS}) and the associated Chern--Simons action leads to spin-3 gravity in three-dimensional flat space-time \cite{Afshar:2013vka, Gonzalez:2013oaa}.

In order to construct Chern--Simons actions for Carroll and Galilei spin-3 algebras, one therefore needs to know whether these algebras can be equipped with an invariant metric. In this respect, it is useful to note that it is not always true that the IW contraction of an algebra equipped with an invariant metric, also admits one. A counter-example is provided by the three-dimensional Carroll and Galilei algebras which both arise as IW contractions of the Poincar\'e algebra, that in three dimensions has an invariant metric. However, only the Carroll algebra admits an invariant metric; the Galilei algebra does not. Naively, one can thus not construct a Chern--Simons action for the Galilei algebra. Fortunately, there exists an extension of the Galilei algebra, the so-called Extended Bargmann algebra \cite{Papageorgiou:2009zc,Papageorgiou:2010ud,Bergshoeff:2016lwr,Hartong:2016yrf}, that can be equipped with an invariant metric and for which a Chern--Simons action can be constructed.

In this section, we will show that similar results hold in the spin-3 case. In particular, we will see that the spin-3 versions of the Carroll algebra admit an invariant metric and that a Chern--Simons action can be straightforwardly constructed. The spin-3 versions of the Galilei algebra do not have an invariant metric but a theorem due to Medina and Revoy \cite{Medina1985} implies that they can be extended to algebras with an invariant metric. We will then explicitly construct these `spin-3 Extended Bargmann' algebras and their associated Chern--Simons actions. In this way, we will obtain spin-3 versions of Carroll gravity \cite{Hartong:2015xda,Bergshoeff:2017btm} and Extended Bargmann gravity \cite{Papageorgiou:2009zc,Papageorgiou:2010ud,Bergshoeff:2016lwr,Hartong:2016yrf}.

We will first treat the case of spin-3 Carroll gravity in section \ref{ssec:CarrollGravity}, while the spin-3 Extended Bargmann gravity case will be discussed in section \ref{ssec:ExtBargGravity}. In both cases, we will also study the equations of motion, at the linearized level. This will allow us to interpret the Chern-Simons actions for these theories as suitable spin-3 generalizations of the actions of Carroll and Extended Bargmann gravity, in a first order formulation. In particular, this linearized analysis will show that some of the gauge fields appearing in these actions can be interpreted as generalized vielbeine, while others can be viewed as generalized spin connections. The latter in particular appear only algebraically in the equations of motion and are therefore dependent fields that can be expressed in terms of other fields. We will give these expressions. In some cases, we will see that not all spin connection components become dependent. We will argue that the remaining independent spin connection components can be viewed as Lagrange multipliers that implement certain constraints on the geometry. For simplicity, we will restrict ourselves to Carroll and Galilei spin-3 gravity theories. The analysis provided here can be straightforwardly extended to include a cosmological constant.

\subsection{Spin-3 Carroll Gravity}
\label{ssec:CarrollGravity}

There are four distinct ways of contracting $\mathfrak{hs}_3\mathfrak{(A)dS}$, such that a spin-3 algebra whose spin-2 part coincides with the Carroll algebra is obtained. These four ways correspond to combining the contraction procedures 1 and 5, 1 and 6, 2 and 5 or 2 and 6 of table \ref{tab:contr} respectively. We will denote the resulting algebras as $\mathfrak{hs}_3\mathfrak{car1}$, $\mathfrak{hs}_3\mathfrak{car2}$, $\mathfrak{hs}_3\mathfrak{car3}$ and $\mathfrak{hs}_3\mathfrak{car4}$. Their structure constants are summarized in table \ref{tab:hscar}. Note that $\mathfrak{hs}_3\mathfrak{car3}$ and $\mathfrak{hs}_3\mathfrak{car4}$ each come in two versions, since we apply the IW contraction procedures to AdS and dS simultaneously. These versions differ in the signs of some of their structure constants, as can be seen from table \ref{tab:hscar}. The existence of these different versions when applying the contraction procedures 2 and 5 (or 2 and 6) stems from the fact that the combination of these contraction procedures leads to different algebras, depending on whether one starts from $\mathfrak{hs}_3\mathfrak{AdS}$ or from $\mathfrak{hs}_3\mathfrak{dS}$. By contrast, applying contraction procedures 1 and 5 (or 1 and 6) on $\mathfrak{hs}_3\mathfrak{AdS}$ and $\mathfrak{hs}_3\mathfrak{dS}$ leads to the same result, namely $\mathfrak{hs}_3\mathfrak{car1}$ (or $\mathfrak{hs}_3\mathfrak{car2}$).

By examining the structure constants of table \ref{tab:hscar}, one can see that $\mathfrak{hs}_3\mathfrak{car1}$ ($\mathfrak{hs}_3\mathfrak{car2}$) and $\mathfrak{hs}_3\mathfrak{car3}$ ($\mathfrak{hs}_3\mathfrak{car4}$) are related via the following interchange of generators
\begin{equation}
  \Ht^a \leftrightarrow \Jt^a \,, \qquad \qquad \Pt^{ab} \leftrightarrow \Gt^{ab}
\end{equation}
plus potentially some sign changes in structure constants, as mentioned in the previous paragraph. The structure of the Chern--Simons theories will therefore be very similar for $\mathfrak{hs}_3\mathfrak{car1}$ ($\mathfrak{hs}_3\mathfrak{car2}$) and $\mathfrak{hs}_3\mathfrak{car3}$ ($\mathfrak{hs}_3\mathfrak{car4}$). In the following, we will therefore restrict ourselves to $\mathfrak{hs}_3\mathfrak{car1}$ and $\mathfrak{hs}_3\mathfrak{car2}$. We will now discuss the Chern--Simons theories for these two cases in turn.

\subsubsection{Chern--Simons Theory for $\mathfrak{hs}_3\mathfrak{car1}$}
\label{sssec:hscar1}

The commutation relations of $\mathfrak{hs}_3\mathfrak{car1}$ are summarized in the first column of table \ref{tab:hscar}. This algebra admits the following invariant metric
\begin{alignat}{2}
  \label{eq:invmetrichscar1}
  & \langle \Ht\,, \Jt \rangle = - 1 \,, \qquad \qquad & & \langle\Pt_a \,, \Gt_b\rangle = \delta_{ab} \,, \nonumber \\
  & \langle \Ht_a\,, \Jt_b \rangle = - \delta_{ab} \,, \qquad \qquad & & \langle\Pt_{ab} \,, \Gt_{cd} \rangle = \delta_{a(c} \delta_{d)b} - \frac23 \delta_{ab} \delta_{cd} \,.
\end{alignat}
This can be obtained either by direct computation or by applying the contraction procedures 1 and 5 of table \ref{tab:contr} on the invariant metric (\ref{eq:bilformAdSdS}) of $\mathfrak{hs}_3\mathfrak{(A)dS}$ (suitably rescaled with the contraction parameters). Using the commutation relations of $\mathfrak{hs}_3\mathfrak{car1}$ and the invariant metric (\ref{eq:invmetrichscar1}), the Chern--Simons action (\ref{eq:CSaction}) and its equations of motion can be explicitly written down. Here, we will be interested in studying the action and equations of motion, linearized around a flat background solution\footnote{For fields in this flat background solution, the curved $\mu$ index becomes equivalent to a flat one. In the following, we will therefore denote the time-like and spatial values of the $\mu$ index by 0 and $a$. The $a$ index can moreover be freely raised and lowered using a Kronecker delta. We will often raise or lower spatial $a$ indices on field components (even if it leads to equations with non-matching index positions on the left- and right-hand-sides), to make more clear which field components are being meant.} given by
\begin{equation}
  \label{eq:flatsol}
  \bar{A}_\mu = \delta_\mu^0 \, \Ht + \delta_\mu^a \, \Pt_a \,.
\end{equation}
We will therefore assume that the gauge field is given by this background solution $\bar{A}_\mu$, plus fluctuations around this background
\begin{equation}
  \label{eq:gaugefieldlin}
  A_\mu = \left(\delta_\mu^0 + \tau_\mu \right) \, \Ht + \left( \delta_\mu^a + e_\mu{}^a \right) \, \Pt_a + \omega_\mu \, \Jt + B_\mu{}^a \, \Gt_a + \tau_\mu{}^a \, \Ht_a + e_\mu{}^{ab} \, \Pt_{ab} + \omega_\mu{}^a \, \Jt_a + B_\mu{}^{ab} \, \Gt_{ab} \,.
\end{equation}
Here, $\tau_\mu$ can be interpreted as a linearized time-like vielbein, $e_\mu{}^a$ as a linearized spatial vielbein, while $\omega_\mu$ and $B_\mu{}^a$ can be viewed as linearized spin connections for spatial rotations and boosts respectively. Similarly, $\tau_\mu{}^a$, $e_\mu{}^{ab}$, $\omega_\mu{}^a$ and $B_\mu{}^{ab}$ can be interpreted as spin-3 versions of these linearized vielbeine and spin connections.

Using the expansion (\ref{eq:gaugefieldlin}) in the Chern--Simons action (\ref{eq:CSaction}) and keeping only the terms quadratic in the fluctuations, one finds the following linearized action:
\begin{align}
  \label{eq:linScar1}
  S_{\mathfrak{hs}_3\mathfrak{car1}} &= \int\, d^3 x \, \epsilon^{\mu\nu\rho} \Big( -2 \tau_\mu \partial_\nu \omega_\rho + 2 e_\mu{}^a \partial_\nu B_{\rho}{}^a - 2 \tau_\mu{}^a \partial_\nu \omega_\rho{}^a + 4 e_\mu{}^{ab} \partial_\nu B_{\rho}{}^{ab} - \frac43 e_{\mu}{}^{aa} \partial_\nu B_\rho{}^{bb} \nonumber \\
  & \qquad \qquad \qquad \ \ - \delta_\mu^0 \omega_\nu{}^a \omega_\rho{}^b \epsilon_{ab} - 2 \delta_\mu^a \omega_\nu B_\rho{}^b \epsilon_{ab} - 4 \delta_\mu^a \omega_\nu{}^c B_\rho{}^{cb} \epsilon_{ab} \Big) \,.
\end{align}
The linearized equations of motion corresponding to this action are given by
\begin{align}
  \label{eq:lineomscar1}
  & 0 = R_{\mu\nu}(\Ht) \equiv  \partial_\mu \tau_\nu - \partial_\nu \tau_\mu - \delta_\mu^a B_\nu{}^b \epsilon_{ab} + \delta_\nu^a B_\mu{}^b \epsilon_{ab} \,, \nonumber \\
  & 0 = R_{\mu\nu}(\Pt^a) \equiv  \partial_\mu e_\nu{}^a - \partial_\nu e_\mu{}^a + \epsilon^{ab} \delta_\mu^b \omega_\nu - \epsilon^{ab} \delta_\nu^b \omega_\mu \,, \nonumber \\
  & 0 = R_{\mu\nu}(\Jt) \equiv  \partial_\mu \omega_\nu - \partial_\nu \omega_\mu \,, \nonumber \\
  & 0 = R_{\mu\nu}(\Gt^a) \equiv  \partial_\mu B_\nu{}^a - \partial_\nu B_\mu{}^a \,, \nonumber \\
  & 0 = R_{\mu\nu}(\Ht^a) \equiv  \partial_\mu \tau_\nu{}^a - \partial_\nu \tau_\mu{}^a - \epsilon^{ab} \delta_\mu^0 \omega_\nu{}^b + \epsilon^{ab} \delta_\nu^0 \omega_\mu{}^b - 2 \delta_\mu^b B_\nu{}^{ac} \epsilon_{bc} + 2 \delta_\nu^b B_\mu{}^{ac} \epsilon_{bc} \,, \nonumber \\
  & 0 = R_{\mu\nu}(\Pt^{ab}) \equiv \partial_\mu e_\nu{}^{ab} - \partial_\nu e_\mu{}^{ab} + \frac12 \delta_\mu^c \omega_\nu{}^{(a} \epsilon^{b)c} - \frac12 \delta_\nu^c \omega_\mu{}^{(a} \epsilon^{b)c} - \delta_\mu^c \omega_\nu{}^d \epsilon_{cd} \delta^{ab} + \delta_\nu^c \omega_\mu{}^d \epsilon_{cd} \delta^{ab} \,, \nonumber \\
  & 0 = R_{\mu\nu}(\Jt^a) \equiv \partial_\mu \omega_\nu{}^a - \partial_\nu \omega_\mu{}^a \,, \nonumber \\
  & 0 = R_{\mu\nu}(\Gt^{ab}) \equiv \partial_\mu B_\nu{}^{ab} - \partial_\nu B_\mu{}^{ab} \,.
\end{align}
The equations
\begin{equation}
  R_{\mu\nu}(\Ht) = 0 \,, \quad R_{\mu\nu}(\Pt^a) = 0 \,, \quad R_{\mu\nu}(\Ht^a) = 0 \,, \quad R_{\mu\nu}(\Pt^{ab}) = 0
\end{equation}
 contain the spin connections $\omega_\mu$, $B_\mu{}^a$, $\omega_\mu{}^a$ and $B_\mu{}^{ab}$ only in an algebraic way. These equations can thus be solved to yield expressions for some of the spin connection components in terms of the vielbeine and their derivatives.

Let us first see how this works for the spin-2 spin connections $\omega_\mu$ and $B_\mu{}^a$. The equation $R_{0a}(\Ht) = 0$ can be straightforwardly solved for $B_0{}^a$:
\begin{equation}
  B_0{}^a = \epsilon^{ab} \left( \partial_0 \tau_b - \partial_b \tau_0 \right) \,.
\end{equation}
Similarly, the equation $R_{ab}(\Ht) = 0$ (or equivalently $\epsilon^{ab} R_{ab}(\Ht) = 0$) can be solved for $B_c{}^c$ (the spatial trace of $B_\mu{}^a$):
\begin{equation}
  B_c{}^c = \frac12 \epsilon^{ab} \left(\partial_a \tau_b - \partial_b \tau_a \right) \,.
\end{equation}
From $R_{ab}(\Pt^c) = 0$ (or equivalently $\epsilon^{ab} R_{ab}(\Pt^c) = 0$) one finds the spatial part of $\omega_\mu$:
\begin{equation}
  \omega_a = \frac12 \epsilon^{bc} \left( \partial_b e_{ca} - \partial_c e_{ba} \right) \,.
\end{equation}
Finally, let us consider the equation $R_{0a}(\Pt_b) = 0$. The anti-symmetric part of this equation $\epsilon^{ab} R_{0a}(\Pt_b) = 0$ can be solved for the time-like part of $\omega_\mu$:
\begin{equation}
  \omega_0 = \frac12 \epsilon^{ab} \left(\partial_a e_{0b} - \partial_0 e_{ab} \right) \,.
\end{equation}
The symmetric part $R_{0(a}(\Pt_{b)}) = 0$ does not contain any spin connection and can be viewed as a constraint on the geometry
\begin{equation}
  \label{eq:constraintcar1}
  \partial_0 e_{(ab)} - \partial_{(a} e_{|0|b)} = 0 \,.
\end{equation}
In summary, we find that $R_{\mu\nu}(\Ht) = 0$ and $R_{\mu\nu}(\Pt^a) = 0$ lead to the constraint (\ref{eq:constraintcar1}) as well as the following solutions for $\omega_\mu$ and $B_\mu{}^a$
\begin{align}
  \label{eq:solspin2conncar1}
  \omega_\mu &= \frac12 \delta_\mu^0 \, \epsilon^{ab} \left(\partial_a e_{0b} - \partial_0 e_{ab} \right) + \frac12 \delta_\mu^a\, \epsilon^{bc} \left( \partial_b e_{ca} - \partial_c e_{ba} \right) \,, \nonumber \\
  B_\mu{}^a &= \delta_\mu^0 \, \epsilon^{ab} \left( \partial_0 \tau_b - \partial_b \tau_0 \right) + \frac14 \delta_\mu^a \, \epsilon^{bc} \left (\partial_b \tau_c - \partial_c \tau_b \right) + \delta_\mu^b \, \tilde{B}_b{}^a \,,
\end{align}
where $\tilde{B}_b{}^a$ is an undetermined traceless tensor. The boost connection $B_\mu{}^a$ is thus not fully determined in terms of $\tau_\mu$ and $e_\mu{}^a$.

A similar reasoning allows one to solve for certain components of the spin-3 connections $\omega_\mu{}^a$ and $B_\mu{}^{ab}$. In particular, the equation $R_{ab}(\Ht^c) = 0$ can be solved for $B_d{}^{da}$, a spatial trace of $B_\mu{}^{ab}$:
\begin{equation}
  B_d{}^{da} = \frac14 \epsilon^{bc} \left(\partial_b \tau_c{}^a - \partial_c \tau_b{}^a\right) \,.
\end{equation}
The equation $R_{ab}(\Pt^{cd}) = 0$ can be solved for the symmetric, spatial part of $\omega_\mu{}^a$:
\begin{equation}
  \omega^{(ab)} = \epsilon^{cd} \left(\partial_c e_d{}^{ab} - \partial_d e_c{}^{ab}\right) - \frac13 \delta^{ab} \epsilon^{cd} \left(\partial_c e_d{}^{ee} - \partial_d e_c{}^{ee} \right) \,.
\end{equation}
The anti-symmetric, spatial part of $\omega_\mu{}^a$ can be found from $R_{0a}(\Ht^a) = 0$:
\begin{equation}
  \epsilon^{ab} \omega_{ab} = \partial_0 \tau_a{}^a - \partial_a \tau_0{}^a \,.
\end{equation}
From the other equations contained in $R_{0b}(\Ht^a) = 0$ one then finds
\begin{equation}
  B_0{}^{ab} = \frac14 \epsilon^{(a|c|} \left( \partial_0 \tau_c{}^{b)} - \partial_c \tau_0{}^{b)} \right) + \frac14 \epsilon^{cd} \left( \partial_c e_d{}^{ab} - \partial_d e_c{}^{ab} \right) - \frac16 \delta^{ab} \epsilon^{cd} \left( \partial_c e_d{}^{ee} - \partial_d e_c{}^{ee} \right) \,.
\end{equation}
The equation $R_{0a}(\Pt_{bc}) = 0$ can be divided into a part that is fully symmetric in the indices $a$, $b$, $c$ and a part that is of mixed symmetry:
\begin{equation}
  R_{0a}(\Pt_{bc}) = 0 \qquad \Leftrightarrow \qquad R^{\mathrm{S}}_{0a}(\Pt_{bc}) = 0 \quad \mathrm{and} \quad R^{\mathrm{MS}}_{0a}(\Pt_{bc}) = 0 \,,
\end{equation}
where
\begin{align}
  R^{\mathrm{S}}_{0a}(\Pt_{bc}) & = \frac13 \left(R_{0a}(\Pt_{bc}) + R_{0c}(\Pt_{ab}) + R_{0b}(\Pt_{ca})\right) \,, \nonumber \\
  R^{\mathrm{MS}}_{0a}(\Pt_{bc}) &= \frac13 \left(2 R_{0a}(\Pt_{bc}) - R_{0c}(\Pt_{ab}) - R_{0b}(\Pt_{ca})\right) \,.
\end{align}
The equation $R^{\mathrm{MS}}_{0a}(\Pt_{bc}) = 0$ can be solved for $\omega_0{}^a$, by noting that
\begin{equation}
  R^{\mathrm{MS}}_{0a}(\Pt_{bc}) = 0 \qquad \Leftrightarrow \qquad \epsilon^{ab} R_{0a}(\Pt_{bc}) = 0 \,.
\end{equation}
The solution one finds is given by
\begin{equation}
  \omega_0{}^a = \frac25 \epsilon^{bc} \left( \partial_b e_0{}^{ca} - \partial_0 e_b{}^{ca} \right) \,.
\end{equation}
The fully symmetric part $R^{\mathrm{S}}_{0a}(\Pt_{bc}) = 0$ can not be used to solve for other spin connection components. Rather, it should be viewed as a constraint on the geometry:
\begin{align}
  \label{eq:constraint2car1}
& \partial_0 e_b{}^{ac} - \partial_b e_0{}^{ac} + \partial_0 e_a{}^{bc} - \partial_a e_0{}^{bc} + \partial_0 e_c{}^{ab} - \partial_c e_0{}^{ab} \nonumber \\
&\quad  + \frac25 \delta_{ac} \left( \partial_b e_0{}^{dd} - \partial_d e_0{}^{bd} + \partial_0 e_d{}^{bd} - \partial_0 e_b{}^{dd} \right) \nonumber \\
&\quad + \frac25 \delta_{bc} \left( \partial_a e_0{}^{dd} - \partial_d e_0{}^{ad} + \partial_0 e_d{}^{ad} - \partial_0 e_a{}^{dd} \right) \nonumber \\
&\quad + \frac25 \delta_{ab} \left( \partial_c e_0{}^{dd} - \partial_d e_0{}^{cd} + \partial_0 e_d{}^{cd} - \partial_0 e_c{}^{dd} \right) = 0 \,.
\end{align}
This constraint can be slightly simplified. By contracting it with $\delta^{bc}$, one finds that
\begin{equation}
  \partial_a e_0{}^{bb} - \partial_0 e_a{}^{bb} = 6 \left(\partial_b e_0{}^{ab} - \partial_0 e_b{}^{ab} \right) \,.
\end{equation}
Using this, one finds that (\ref{eq:constraint2car1}) simplifies to
\begin{align}
  \label{eq:constraint2car1v2}
& \partial_0 e_b{}^{ac} - \partial_b e_0{}^{ac} + \partial_0 e_a{}^{bc} - \partial_a e_0{}^{bc} + \partial_0 e_c{}^{ab} - \partial_c e_0{}^{ab}  + \frac13 \delta_{bc} \left( \partial_a e_0{}^{dd} - \partial_0 e_a{}^{dd} \right) \nonumber \\ & \quad + \frac13 \delta_{ac} \left( \partial_b e_0{}^{dd} - \partial_0 e_b{}^{dd} \right) + \frac13 \delta_{ab} \left( \partial_c e_0{}^{dd} - \partial_0 e_c{}^{dd} \right) = 0 \,.
\end{align}
One thus finds for the spin-3 sector, that the equations $R_{\mu\nu}(\Ht^a) = 0$ and $R_{\mu\nu}(\Pt^{ab}) = 0$ lead to the constraint (\ref{eq:constraint2car1v2}) and the following solutions for $\omega_\mu{}^a$ and $B_\mu{}^{ab}$:
\begin{align}
  \label{eq:solspin3conncar1}
  \omega_\mu{}^a &= \frac25 \delta_\mu^0 \epsilon^{bc} \left( \partial_b e_0{}^{ca} - \partial_0 e_b{}^{ca} \right) + \frac12 \delta_\mu^b \Big( \epsilon^{cd} \left( \partial_c e_d{}^{ba} - \partial_d e_c{}^{ba} \right) \nonumber \\ & \qquad - \frac13 \delta_b^a \epsilon^{cd} \left( \partial_c e_d{}^{ee} - \partial_d e_c{}^{ee} \right) + \epsilon_{ba} \left(\partial_0 \tau_c{}^c - \partial_c \tau_0{}^c \right)\Big) \,, \nonumber \\
  B_\mu{}^{ab} &= \frac14 \delta_\mu^0 \left(\epsilon^{(a|c|} \left( \partial_0 \tau_c{}^{b)} - \partial_c \tau_0{}^{b)} \right) + \epsilon^{cd} \left( \partial_c e_d{}^{ab} - \partial_d e_c{}^{ab} \right) - \frac23 \delta^{ab} \epsilon^{cd} \left( \partial_c e_d{}^{ee} - \partial_d e_c{}^{ee} \right) \right) \nonumber \\
& \qquad + \frac{1}{12}  \delta_\mu^{(a} \epsilon^{|de|} \left(\partial_d \tau_e{}^{b)} - \partial_e \tau_d{}^{b)}\right) + \delta_\mu^c \tilde{B}_c{}^{ab} \,,
\end{align}
where $\tilde{B}_c{}^{ab}$ is an arbitrary tensor obeying $\tilde{B}_b{}^{ba} = 0$. As for the spin-2 sector, one thus finds that the spin-3 boost connection $B_\mu{}^{ab}$ can not be fully determined in terms of $\tau_\mu{}^a$ and $e_\mu{}^{ab}$.

It is interesting to see what role the undetermined components $\tilde{B}_b{}^a$ and $\tilde{B}_c{}^{ab}$ play. In particular, one can check how these components appear in the Lagrangian and what their equations of motion are. Upon partial integration in the action (\ref{eq:linScar1}), one finds that the terms in the Lagrangian involving $B_\mu{}^a$ can be written as
\begin{equation}
  \epsilon^{\mu\nu\rho} R_{\mu\nu}(\Pt_a) B_\rho{}^a \,.
\end{equation}
The traceless spatial components $\tilde{B}_b{}^a$ of $B_\rho{}^a$ thus couple to
\begin{equation}
  \epsilon^{cb} R_{0c}(\Pt_a) - \frac12 \delta_a^{b} \epsilon^{cd} R_{0c}(\Pt_d) \,.
\end{equation}
This can however be rewritten as
\begin{equation}
 - \frac12 \epsilon^{cb} R_{0(a}(\Pt_{b)}) \,.
\end{equation}
One thus sees that $\tilde{B}_b{}^a$ acts as a Lagrange multiplier for $R_{0(a}(\Pt_{b)}) = 0$, which led to the constraint (\ref{eq:constraintcar1}). Similarly, one can check that $\tilde{B}_c{}^{ab}$ plays the role of Lagrange multiplier for the constraint (\ref{eq:constraint2car1v2}).

\subsubsection{Chern--Simons Theory for $\mathfrak{hs}_3\mathfrak{car2}$}
\label{sssec:hscar2}

The commutation relations of $\mathfrak{hs}_3\mathfrak{car2}$ can be obtained by successively applying the contraction procedures 1 and 6 of table \ref{tab:contr} and are given in the second column of table \ref{tab:hscar}. One can check, either by explicit computation or by contraction of the invariant metric (\ref{eq:bilformAdSdS}), that $\mathfrak{hs}_3\mathfrak{car2}$ can be equipped with the same invariant metric (\ref{eq:invmetrichscar1}) as $\mathfrak{hs}_3\mathfrak{car1}$. Similarly, it can be checked that (\ref{eq:flatsol}) is a solution of the full non-linear equations of motion of the Chern--Simons theory for $\mathfrak{hs}_3\mathfrak{car2}$. As before, we will study the Chern--Simons action and its equations of motion, linearized around this background. Adopting the linearization ansatz (\ref{eq:gaugefieldlin}), the linearized Chern--Simons action is found to be given by
\begin{align}
  \label{eq:linScar2}
  S_{\mathfrak{hs}_3\mathfrak{car2}} &= \int\, d^3 x \, \epsilon^{\mu\nu\rho} \Big(- 2 \tau_\mu \partial_\nu \omega_\rho + 2 e_\mu{}^a \partial_\nu B_{\rho}{}^a - 2 \tau_\mu{}^a \partial_\nu \omega_\rho{}^a + 4 e_\mu{}^{ab} \partial_\nu B_{\rho}{}^{ab} - \frac43 e_\mu{}^{aa} \partial_\nu B_\rho{}^{bb} \nonumber \\
& \qquad \qquad \qquad + 4 \delta_\mu^0 B_\nu{}^{ac} B_\rho{}^{cb} \epsilon_{ab} - 2 \delta_\mu^a \omega_\nu B_\rho{}^b \epsilon_{ab} - 4 \delta_\mu^a \omega_\nu{}^c B_\rho{}^{cb} \epsilon_{ab} \Big) \,.
\end{align}
The linearized equations of motion derived from this action are
\begin{align}
  \label{eq:lineomscar2}
  & 0 = R_{\mu\nu}(\Ht) \equiv  \partial_\mu \tau_\nu - \partial_\nu \tau_\mu - \delta_\mu^a B_\nu{}^b \epsilon_{ab} + \delta_\nu^a B_\mu{}^b \epsilon_{ab} \,, \nonumber \\
  & 0 = R_{\mu\nu}(\Pt^a) \equiv  \partial_\mu e_\nu{}^a - \partial_\nu e_\mu{}^a + \epsilon^{ab} \delta_\mu^b \omega_\nu - \epsilon^{ab} \delta_\nu^b \omega_\mu \,, \nonumber \\
  & 0 = R_{\mu\nu}(\Jt) \equiv  \partial_\mu \omega_\nu - \partial_\nu \omega_\mu \,, \nonumber \\
  & 0 = R_{\mu\nu}(\Gt^a) \equiv  \partial_\mu B_\nu{}^a - \partial_\nu B_\mu{}^a \,, \nonumber \\
  & 0 = R_{\mu\nu}(\Ht^a) \equiv  \partial_\mu \tau_\nu{}^a - \partial_\nu \tau_\mu{}^a - 2 \delta_\mu^b B_\nu{}^{ac} \epsilon_{bc} + 2 \delta_\nu^b B_\mu{}^{ac} \epsilon_{bc} \,, \nonumber \\
  & 0 = R_{\mu\nu}(\Pt^{ab}) \equiv \partial_\mu e_\nu{}^{ab} - \partial_\nu e_\mu{}^{ab} + \frac12 \delta_\mu^c \omega_\nu{}^{(a} \epsilon^{b)c} - \frac12 \delta_\nu^c \omega_\mu{}^{(a} \epsilon^{b)c} - \delta_\mu^c \omega_\nu{}^d \epsilon_{cd} \delta^{ab} + \delta_\nu^c \omega_\mu{}^d \epsilon_{cd} \delta^{ab}  \nonumber \\
& \qquad \qquad \qquad \quad - \delta_\mu^0 B_\nu{}^{c(a} \epsilon^{b)c} + \delta_\nu^0 B_\mu{}^{c(a} \epsilon^{b)c} \,, \nonumber \\
  & 0 = R_{\mu\nu}(\Jt^a) \equiv \partial_\mu \omega_\nu{}^a - \partial_\nu \omega_\mu{}^a \,, \nonumber \\
  & 0 = R_{\mu\nu}(\Gt^{ab}) \equiv \partial_\mu B_\nu{}^{ab} - \partial_\nu B_\mu{}^{ab} \,.
\end{align}
As in the case of $\mathfrak{hs}_3\mathfrak{car1}$, the equations $R_{\mu\nu}(\Ht) = 0$, $R_{\mu\nu}(\Pt^a) = 0$, $R_{\mu\nu}(\Ht^a) = 0$ and $R_{\mu\nu}(\Pt^{ab}) = 0$ contain the spin connections $\omega_\mu$, $B_\mu{}^a$, $\omega_\mu{}^a$ and $B_\mu{}^{ab}$ algebraically. These equations can thus be used to express some spin connection components in terms of the vielbeine. Since the linearized equations of motion for $\mathfrak{hs}_3\mathfrak{car1}$ and $\mathfrak{hs}_3\mathfrak{car2}$ coincide for the spin-2 sector, the solutions for $\omega_\mu$ and $B_\mu{}^a$ are again given by (\ref{eq:solspin2conncar1}). As before, the traceless spatial part $\tilde{B}_b{}^a$ of $B_\mu{}^a$ acts as a Lagrange multiplier for the constraint (\ref{eq:constraintcar1}).

In the spin-3 sector, one can solve the equations $R_{cd}(\Pt^{ab}) = 0$ for the symmetric, spatial part of $\omega_\mu{}^a$:
\begin{equation}
  \omega^{(ab)} = \epsilon^{cd} \left( \partial_c e_d{}^{ab} - \partial_d e_c{}^{ab} \right) - \frac13 \delta^{ab} \epsilon^{cd} \left( \partial_c e_d{}^{ee} - \partial_d e_c{}^{ee} \right) \,.
\end{equation}
From $R_{0c}(\Pt^{aa}) = 0$, one finds
\begin{equation}
  \omega_0{}^a = \frac13 \epsilon^{ab} \left( \partial_0 e_b{}^{cc} - \partial_b e_0{}^{cc} \right) \,,
\end{equation}
while the remaining equations in $R_{0c}(\Pt^{aa}) = 0$, together with $R_{bc}(\Ht^a) = 0$ lead to
\begin{align}
B_a{}^{bc} &= \frac12 \epsilon^{cd} \left(\partial_d e_0{}^{ab} - \partial_0 e_d{}^{ab} \right)  - \frac12 \delta^{ac} \epsilon^{de} \left( \partial_0 e_d{}^ {eb} - \partial_d e_0{}^{eb} \right) + \frac12 \delta^{bc} \epsilon^{de} \left( \partial_0 e_d{}^{ea} - \partial_d e_0{}^{ea} \right) \nonumber \\
& \quad + \frac14 \delta^{bc} \epsilon^{de} \left( \partial_d \tau_e{}^a - \partial_e \tau_d{}^a \right) + \frac12 \delta^{bc} \epsilon^{ad} \left( \partial_0 e_d{}^{ee} - \partial_d e_0{}^{ee} \right) - \frac13 \delta^{ac} \epsilon^{bd} \left( \partial_0 e_d{}^{ee} - \partial_d e_0{}^{ee} \right) \nonumber \\
& \quad + \frac16 \delta^{ab} \epsilon^{cd} \left( \partial_0 e_d{}^{ee} - \partial_d e_0{}^{ee} \right) \,.
\end{align}
The traceless part of $R_{0b}(\Ht^a) = 0$ can be used to solve for $B_0{}^{ab}$:
\begin{equation}
  B_0{}^{ab} = \frac14 \epsilon^{(a|c|} \left(\partial_0 \tau_c{}^{b)} - \partial_c \tau_0{}^{b)} \right) \,,
\end{equation}
while the trace part $R_{0a}(\Ht^a) = 0$ leads to the constraint
\begin{equation} \label{eq:constrcar2}
  \partial_0 \tau_a{}^a - \partial_a \tau_0{}^a = 0 \,.
\end{equation}
One thus finds
\begin{align}
  \label{eq:solspin3conncar2}
  B_\mu{}^{ab} &= \frac14 \delta_\mu^0 \epsilon^{(a|c|} \left(\partial_0 \tau_c{}^{b)} - \partial_c \tau_0{}^{b)} \right) + \delta_\mu^c \Bigg(\frac12 \epsilon^{ad} \left(\partial_d e_0{}^{bc} - \partial_0 e_d{}^{bc} \right)  - \frac12 \delta^{ac} \epsilon^{de} \left( \partial_0 e_d{}^ {eb} - \partial_d e_0{}^{eb} \right) \nonumber \\ & \quad + \frac12 \delta^{ab} \epsilon^{de} \left( \partial_0 e_d{}^{ec} - \partial_d e_0{}^{ec} \right)  + \frac14 \delta^{ab} \epsilon^{de} \left( \partial_d \tau_e{}^c - \partial_e \tau_d{}^c \right) + \frac12 \delta^{ab} \epsilon^{cd} \left( \partial_0 e_d{}^{ee} - \partial_d e_0{}^{ee} \right)\nonumber \\ & \quad - \frac13 \delta^{ac} \epsilon^{bd} \left( \partial_0 e_d{}^{ee} - \partial_d e_0{}^{ee} \right) + \frac16 \delta^{bc} \epsilon^{ad} \left( \partial_0 e_d{}^{ee} - \partial_d e_0{}^{ee} \right) \Bigg) \,, \nonumber \\
  \omega_\mu{}^a &= \frac13 \delta_\mu^0 \epsilon^{ab} \left( \partial_0 e_b{}^{cc} - \partial_b e_0{}^{cc} \right) + \delta_\mu^a \epsilon^{cd} \Big(\left(\partial_c e_d{}^{ab} - \partial_d e_c{}^{ab} \right) - \frac13 \delta^{ab} \left( \partial_c e_d{}^{ee} - \partial_d e_c{}^{ee} \right)\Big) \nonumber \\ & \quad + \delta_\mu^b \, \epsilon^{ab}\,\tilde{\omega} \,,
\end{align}
where $\tilde{\omega}$ is undetermined. By examining the action (\ref{eq:linScar2}), one can see that $\tilde{\omega}$ plays the role of a Lagrange multiplier for the constraint (\ref{eq:constrcar2}).

\subsection{Spin-3 Galilei and Extended Bargmann Gravity}
\label{ssec:ExtBargGravity}

In the previous section, we have studied Carroll spin-3 algebras, whose spin-2 part corresponds to the Carroll algebra. Using the contraction procedures of table \ref{tab:contr}, one can also obtain non-relativistic spin-3 algebras, that contain the Galilei algebra. As in the Carroll case, there are four distinct ways of doing this, namely by successively applying the contraction procedures 1 and 3, 1 and 4, 2 and 3 or 2 and 4 of table \ref{tab:contr}. We have called the resulting algebras $\mathfrak{hs}_3\mathfrak{gal1}$, $\mathfrak{hs}_3\mathfrak{gal2}$, $\mathfrak{hs}_3\mathfrak{gal3}$ and $\mathfrak{hs}_3\mathfrak{gal4}$ respectively and summarized their commutation relations in table \ref{tab:hsgal}. As in the Carroll case, $\mathfrak{hs}_3\mathfrak{gal3}$ and $\mathfrak{hs}_3\mathfrak{gal4}$ each come in two different versions, depending on whether one applies the combination of contraction procedures on $\mathfrak{hs}_3\mathfrak{AdS}$ or $\mathfrak{hs}_3\mathfrak{dS}$. They are again structurally similar to $\mathfrak{hs}_3\mathfrak{gal1}$ and $\mathfrak{hs}_3\mathfrak{gal2}$. We will therefore restrict our discussion here to these two cases.

In contrast to the spin-3 Carroll algebras, whose invariant metrics arose from applying the relevant contraction procedures on (\ref{eq:bilformAdSdS}), a similar reasoning for the spin-3 Galilei algebras leads to degenerate bilinear forms. One can in fact show by direct computation that they can not be equipped with a nondegenerate symmetric invariant bilinear form\footnote{This is even true when one allows nontrivial central extensions. One algebra admits no nontrivial central extensions (the second cohomology group is trivial), whereas the other does admit three nontrivial extensions of which no combination of them can be used to define an invariant metric.}. It could be interesting to investigate these algebras, given explicitly in table \ref{tab:hsgal}, and their degenerate bilinear forms. For the spin-2 case, this has been done in \cite{Bergshoeff:2017btm}. Due to the degeneracy of the bilinear form, some of the fields appear without kinetic term  in the action (see equation \eqref{eq:CSaction}) and are therefore not dynamical. In the spin-2 case, one can nevertheless interpret these non-dynamical fields as Lagrange multipliers for geometrical constraints, similarly to what happens in the Carroll cases of the previous section. Although it would be interesting to see whether similar results hold for the higher spin case, we will not do this here and instead we will look at Chern-Simons theories where each field has a kinetic term. These can not be based on the spin-3 Galilei algebras, but interestingly, a theorem due to Medina and Revoy \cite{Medina1985} implies that these algebras can be extended to algebras that admit an invariant metric, i.e. a nondegenerate ad-invariant symmetric bilinear form. Remarkably, in this way  one ends up with a spin-3 version of the Extended Bargmann algebra. This procedure can be stated  as follows \cite{Medina1985,FigueroaO'Farrill:1995cy}.
\begin{theorem}
  Consider a Lie algebra that is a semi-direct sum of Lie algebras $\mathfrak{g}$ (with generators $\Gt_i$) and $\mathfrak{h}$ (with generators $\Ht_\alpha$), whose commutation relation are given by
  \begin{align} \label{eq:commsmr}
    \left[\, \Gt_i \comma \Gt_j \, \right] &= f_{ij}{}^k \Gt_k \qquad \left[\, \Ht_\alpha \comma \Gt_i \, \right] = f_{\alpha i}{}^j \Gt_j \qquad \left[\, \Ht_\alpha \comma \Ht_\beta \, \right] = f_{\alpha \beta}{}^\gamma \Ht_\gamma \,.
  \end{align}
Suppose furthermore that $\mathfrak{g}$ is equipped with an  invariant metric $\langle \cdot \,, \cdot \rangle_{\mathfrak{g}}$
\begin{align}
  \label{eq:trmr}
  \langle\Gt_i\,, \Gt_j\rangle_{\mathfrak{g}} = g_{ij} \,,
\end{align}
that is left invariant by the action of $\mathfrak{h}$ on $\mathfrak{g}$, given by the second commutation relation of (\ref{eq:commsmr}). Denote the dual of $\mathfrak{h}$ by $\mathfrak{h}^*$ (with generators $\Ht^\alpha$).

Then, $\mathfrak{g} + \mathfrak{h} + \mathfrak{h}^*$ (where $+$ denotes a direct sum as vector spaces) forms a Lie algebra $\mathfrak{d}$ with commutation relations\begin{alignat}{2} \label{eq:commsmr2}
  &[\,\Gt_i \comma \Gt_j \,] = f\indices{_{ij}^k} \Gt_k + f_{ij\alpha} \Ht^\alpha \qquad &  &[\, \Ht_\alpha \comma \Ht_\beta \,] = f\indices{_{\alpha\beta}^\gamma} \Ht_\gamma  \nonumber \\
  &[\,\Ht_\alpha \comma \Gt_i\,] = f\indices{_{\alpha i}^j} \Gt_j  & &[\, \Ht_\alpha \comma \Ht^\beta \,] = - f\indices{_{\alpha \gamma}^\beta} \Ht^\gamma  \nonumber \\
  &[\,\Ht^\alpha \comma \Gt_i \,] = 0   & &[\,\Ht^\alpha \comma \Ht^\beta \,] = 0 \,,
\end{alignat}
where $f_{ij\alpha} = f_{\alpha i}{}^k g_{kj}$. This Lie algebra $\mathfrak{d}$, called the double extension of $\mathfrak{g}$ by $\mathfrak{h}$, moreover admits an invariant metric $\langle \cdot \,, \cdot \rangle_{\mathfrak{d}}$ whose non-zero components are given by
\begin{equation}
  \label{eq:trmr2}
  \langle \Gt_i\,, \Gt_j\rangle_{\mathfrak{d}} = g_{ij} \,, \qquad \langle\Ht_\alpha\,,\Ht_\beta\rangle_{\mathfrak{d}} = h_{\alpha\beta} \,, \qquad \langle \Ht_\alpha\,,\Ht^\beta \rangle_{\mathfrak{d}} = \delta_\alpha^\beta \,,
\end{equation}
where $h_{\alpha\beta}$ represents a possibly degenerate symmetric invariant bilinear form on $\mathfrak{h}$.
\end{theorem}
This theorem can be applied to the ordinary Galilei algebra in three dimensions and yields the so-called Extended Bargmann algebra \cite{Papageorgiou:2009zc,Papageorgiou:2010ud,Bergshoeff:2016lwr,Hartong:2016yrf}, that extends the Galilei algebra with two central extensions. Applying the theorem to $\mathfrak{hs}_3 \mathfrak{gal1}$ and $\mathfrak{hs}_3\mathfrak{gal2}$ yields two spin-3 algebras, that we will denote, in hindsight,  by $\mathfrak{hs}_3 \mathfrak{ebarg1}$ and $\mathfrak{hs}_3 \mathfrak{ebarg2}$ (since they have an Extended Bargmann spin-2 subalgebra).

The algebra $\mathfrak{hs}_3\mathfrak{ebarg1}$ can be obtained by applying the Medina-Revoy theorem to $\mathfrak{hs}_3\mathfrak{gal1}$. Indeed, with the choices $\mathfrak{g} = \{\Pt_a, \Gt_a, \Pt_{ab}, \Gt_{ab}\}$, $\mathfrak{h} = \{\Ht, \Jt, \Ht_a, \Jt_a\}$ and
\begin{equation}
  \label{eq:trggal1}
  \langle \Pt_a \,, \Gt_b\rangle_{\mathfrak{g}} = \delta_{ab} \,, \qquad \langle\Pt_{ab}\,, \Gt_{cd}\rangle_{\mathfrak{g}} = \delta_{a(c} \delta_{d)b} - \frac23 \delta_{ab} \delta_{cd} \,,
\end{equation}
the assumptions of the theorem are fulfilled and the algebra $\mathfrak{hs}_3\mathfrak{ebarg1}$ can be constructed. Denoting the generators of $\mathfrak{h}^*$ by $\{\Ht^*, \Jt^*, \Ht_a^*, \Jt_a^*\}$, the commutation relations of $\mathfrak{hs}_3\mathfrak{ebarg1}$ are given in table \ref{tab:ebarg1}.
\begin{table}[]
  \centering
$
\begin{array}{l r r l r r r}
\toprule %
                                               & \mathfrak{hs}_3\mathfrak{ebarg1}                  & \quad \quad\quad\quad\quad     &                                              & \mathfrak{hs}_3\mathfrak{ebarg2}     \\ \midrule  
  \left[\,\Jt  \comma \Gt_{a} \,\right]        & \epsilon_{am}  \Gt_{m}                            &                                 & \left[\,\Jt  \comma \Gt_{a} \,\right]        & \epsilon_{am}  \Gt_{m}               \\           
  \left[\, \Jt \comma \Pt_{a} \,\right]        & \epsilon_{am}  \Pt_{m}                            &                                 & \left[\, \Jt \comma \Pt_{a} \,\right]        & \epsilon_{am}  \Pt_{m}               \\           
  \left[\, \Gt_{a} \comma \Ht \,\right]        & -\epsilon_{am}  \Pt_{m}                           &                                 & \left[\, \Gt_{a} \comma \Ht \,\right]        & -\epsilon_{am}  \Pt_{m}              \\           
    \left[\, \Gt_a \comma \Gt_b \, \right]     & \epsilon_{ab} \Ht^*                               &                                 & \left[\, \Gt_a\comma \Gt_b\,\right]          & \epsilon_{ab} \Ht^*                  \\           
  \left[\, \Pt_a \comma \Gt_b \, \right]       & \epsilon_{ab} \Jt^*                               &                                 & \left[\, \Gt_a\comma \Pt_b\,\right]          & \epsilon_{ab} \Jt^*                  \\\midrule   
  \left[\, \Jt \comma \Jt_{a} \,\right]        & \epsilon_{am} \Jt_{m}                             &                                 & \left[\, \Jt \comma \Jt_{a} \,\right]        & \epsilon_{am} \Jt_{m}                \\           
  \left[\, \Jt \comma \Gt_{ab} \,\right]       & -\epsilon_{m(a}\Gt_{b)m}                          &                                 & \left[\, \Jt \comma \Gt_{ab} \,\right]       & -\epsilon_{m(a}\Gt_{b)m}             \\           
  \left[\, \Jt \comma \Ht_{a} \,\right]        & \epsilon_{am} \Ht_{m}                             &                                 & \left[\, \Jt \comma \Ht_{a} \,\right]        & \epsilon_{am} \Ht_{m}                \\           
  \left[\, \Jt \comma \Pt_{ab} \,\right]       & -\epsilon_{m(a}\Pt_{b)m}                          &                                 & \left[\, \Jt \comma \Pt_{ab} \,\right]       & -\epsilon_{m(a}\Pt_{b)m}             \\           
  \left[\, \Gt_{a} \comma \Jt_{b}  \,\right]   & -(\epsilon_{am} \Gt_{bm}+ \epsilon_{ab} \Gt_{mm}) &                                 & \left[\, \Gt_{a} \comma \Gt_{bc}  \,\right]  & -\epsilon_{a(b}\Jt_{c)}              \\          
  \left[\, \Gt_{a} \comma \Ht_{b}  \,\right]   & -(\epsilon_{am} \Pt_{bm}+ \epsilon_{ab} \Pt_{mm}) &                                 & \left[\, \Gt_{a} \comma \Pt_{bc}  \,\right]  & -\epsilon_{a(b}\Ht_{c)}              \\          
  \left[\, \Ht \comma \Jt_{a} \,\right]        & \epsilon_{am} \Ht_{m}                             &                                 & \left[\, \Ht \comma \Jt_{a} \,\right]        & \epsilon_{am} \Ht_{m}                \\           
  \left[\, \Ht \comma \Gt_{ab} \,\right]       & -\epsilon_{m(a}\Pt_{b)m}                          &                                 & \left[\, \Ht \comma \Gt_{ab} \,\right]       & -\epsilon_{m(a}\Pt_{b)m}             \\           
  \left[\, \Pt_{a} \comma \Jt_{b}  \,\right]   & -(\epsilon_{am} \Pt_{bm}+ \epsilon_{ab} \Pt_{mm})  &                                 & \left[\, \Pt_{a} \comma \Gt_{bc}  \,\right]  & -\epsilon_{a(b}\Ht_{c)}              \\ \midrule 
  \left[\, \Jt_{a} \comma \Jt_{b} \,\right]    & \epsilon_{ab} \Jt                                 &                                 & \left[\, \Jt_{a} \comma \Gt_{bc} \,\right]   & \delta_{a(b}\epsilon_{c)m} \Gt_{m}   \\           
  \left[\, \Jt_{a} \comma \Gt_{bc} \,\right]   & \delta_{a(b}\epsilon_{c)m} \Gt_{m}                &                                 & \left[\, \Jt_{a} \comma \Pt_{bc} \,\right]   & \delta_{a(b}\epsilon_{c)m} \Pt_{m}   \\           
  \left[\, \Jt_{a} \comma \Ht_{b} \,\right]    & \epsilon_{ab} \Ht                                 &                                 & \left[\, \Gt_{ab} \comma \Gt_{cd}  \,\right] & \delta_{(a(c}\epsilon_{d)b)}\Jt      \\           
  \left[\, \Jt_{a} \comma \Pt_{bc} \,\right]   & \delta_{a(b}\epsilon_{c)m} \Pt_{m}                &                                 & \left[\, \Gt_{ab} \comma \Ht_{c} \,\right]   & - \delta_{c(a}\epsilon_{b)m} \Pt_{m} \\           
  \left[\, \Gt_{ab} \comma \Ht_{c} \,\right]   & - \delta_{c(a}\epsilon_{b)m} \Pt_{m}              &                                 & \left[\, \Gt_{ab} \comma \Pt_{cd}  \,\right] & \delta_{(a(c}\epsilon_{d)b)}\Ht      \\ \midrule  
  \left[\, \Gt_{ab} \comma \Gt_{cd}  \,\right] & \epsilon_{(a(c}\delta_{d)b)} \Ht^{*}              &                                 & \left[\, \Gt_a\comma \Jt_b\,\right]          & -\epsilon_{am} \Pt^*_{mb}            \\           
  \left[\, \Pt_{ab} \comma \Gt_{cd}  \,\right] & \epsilon_{(a(c}\delta_{d)b)} \Jt^{*}              &                                 & \left[\, \Gt_a\comma \Ht_b\,\right]          & - \epsilon_{am} \Gt^*_{mb}           \\           
  \left[\, \Pt_{a} \comma \Gt_{bc}  \,\right]  & \epsilon_{a(b} \Jt_{c)}^{*}                       &                                 & \left[\, \Pt_a\comma \Jt_b\,\right]          & - \epsilon_{am} \Gt^*_{mb}           \\           
  \left[\, \Gt_{a} \comma \Gt_{bc}  \,\right]  & \epsilon_{a(b} \Ht_{c)}^{*}                       &                                 & \left[\, \Jt_a\comma \Jt_b\,\right]          & - \epsilon_{ab} \Ht^*                \\           
  \left[\, \Gt_{a} \comma \Pt_{bc}  \,\right]  & \epsilon_{a(b} \Jt_{c)}^{*}                       &                                 & \left[\, \Jt_a\comma \Ht_b\,\right]          & - \epsilon_{ab} \Jt^*                \\ \midrule   
  \left[\, \Jt\comma \Ht^*_a\,\right]          & \epsilon_{am} \Ht^*_m                             &                                 & \left[\,\Jt\comma \Pt^*_{ab}\,\right]        & -\epsilon_{m(a} \Pt^*_{b)m}          \\           
  \left[\, \Jt\comma \Jt^*_a \,\right]         & \epsilon_{am} \Jt^*_m                             &                                 & \left[\,\Jt\comma \Gt^*_{ab}\,\right]        & -\epsilon_{m(a} \Gt^*_{b)m}          \\           
  \left[\, \Ht\comma \Ht^*_a \,\right]         & \epsilon_{am} \Jt^*_m                             &                                 & \left[\, \Ht \comma \Pt^*_{ab}\,\right]      & -\epsilon_{m(a} \Gt_{b)m}^*          \\           
  \left[\, \Jt_a\comma \Jt^* \,\right]         & - \epsilon_{am} \Jt^*_m                           &                                 & \left[\,\Gt_{ab}\comma\Jt^*\,\right]         & - \epsilon_{m(a} \Gt^*_{b)m}         \\           
  \left[\,\Jt_a\comma \Ht^*\,\right]           & - \epsilon_{am} \Ht^*_m                           &                                 & \left[\,\Gt_{ab}\comma\Ht^*\,\right]         & - \epsilon_{m(a} \Pt^*_{b)m}         \\           
  \left[\, \Jt_a\comma \Jt^*_b \,\right]       & \epsilon_{ab} \Jt^*                               &                                 & \left[\,\Gt_{ab}\comma \Gt^*_{cd}\,\right]   & \epsilon_{(a(c} \delta_{d)b)}  \Jt^* \\           
  \left[\,\Jt_a\comma \Ht^*_b\,\right]         & \epsilon_{ab} \Ht^*                               &                                 & \left[\,\Gt_{ab}\comma \Pt^*_{cd}\,\right]   & \epsilon_{(a(c} \delta_{d)b)} \Ht^*  \\           
  \left[\,\Ht_a\comma \Ht^*\,\right]           & - \epsilon_{am} \Jt^*_m                           &                                 & \left[\,\Pt_{ab}\comma\Ht^*\,\right]         & - \epsilon_{m(a} \Gt^*_{b)m}         \\           
  \left[\, \Ht_a\comma \Ht^*_b \,\right]       & \epsilon_{ab} \Jt^*                               &                                 & \left[\,\Pt_{ab}\comma \Pt^*_{cd}\,\right]   & \epsilon_{(a(c} \delta_{d)b)} \Jt^*  \\           
\bottomrule
\end{array}
$
\caption{Nonzero commutators of $\mathfrak{hs}_3\mathfrak{ebarg1}$ and $\mathfrak{hs}_3\mathfrak{ebarg2}$. This algebras admit an invariant metric, given by equation \eqref{eq:trebarg1} and \eqref{eq:trebarg2}, respectively.}
\label{tab:ebarg1}
\end{table}
The invariant metric of $\mathfrak{hs}_3\mathfrak{ebarg1}$ is explicitly given by
\begin{alignat}{2} \label{eq:trebarg1}
 \langle \Pt_a\,, \Gt_b\rangle                  & =  \delta_{ab} \,, \qquad                                                        &
\langle \Pt_{ab}\,, \Gt_{cd}\rangle              & = \delta_{a(c} \delta_{d)b} - \frac23 \delta_{ab}\delta_{cd} \,, \nonumber                                                                                                                              \\
  \langle \Ht\,, \Ht^*\rangle                    & = 1 \,, \qquad                                                                   & \langle \Jt\,, \Jt^*\rangle     & = 1 \,, \nonumber                                                                                         \\
  \langle \Ht_a\,, \Ht^*_b\rangle                & = \delta_{ab} \,, \qquad                                                         & \langle \Jt_a\,, \Jt^*_b\rangle & = \delta_{ab} \,.
\end{alignat}
Similarly, starting from $\mathfrak{hs}_3\mathfrak{gal2}$ and using the Medina-Revoy theorem with the choices $\mathfrak{g} = \{\Pt_a, \Gt_a, \Ht_a, \Jt_a\}$, $\mathfrak{h} = \{\Ht, \Jt, \Pt_{ab},\Gt_{ab}\}$ and
\begin{equation}
  \langle \Pt_a \,, \Gt_b \rangle_{\mathfrak{g}} = \delta_{ab} \,, \qquad \langle \Ht_a\,, \Jt_b \rangle_{\mathfrak{g}} = - \delta_{ab} \,,
\end{equation}
the algebra $\mathfrak{hs}_3 \mathfrak{ebarg2}$ can be constructed. Denoting the generators of $\mathfrak{h}^*$ by $\{\Ht^*, \Jt^*, \Pt^*_{ab}, \Gt^*_{ab}\}$, its commutation relations are given in table \ref{tab:ebarg1}. 

This algebra admits the following invariant metric
\begin{alignat}{2} \label{eq:trebarg2}
  \langle \Pt_a, \Gt_b\rangle                  & =  \delta_{ab} \,, \qquad                                                                            &
\langle \Ht_{a}, \Jt_{b}\rangle                & = - \delta_{ab} \,, \nonumber                                                                                                                                                                                                                     \\
  \langle \Ht, \Ht^*\rangle                    & = 1 \,, \qquad                                                                                       & \langle \Jt, \Jt^*\rangle           & = 1 \,, \nonumber                                                                                         \\
  \langle \Pt_{ab}, \Pt^*_{cd}\rangle          & =  \delta_{a(c} \delta_{d)b} \,, \qquad                                                           & \langle \Gt_{ab}, \Gt^*_{cd}\rangle & =  \delta_{a(c}\delta_{d)b} \,.
\end{alignat}
Note that for both $\mathfrak{hs}_3\mathfrak{ebarg1}$ and $\mathfrak{hs}_3\mathfrak{ebarg2}$ the generators $\{\Ht, \Jt, \Pt_a, \Gt_a, \Ht^*, \Jt^*\}$ form a subalgebra that coincides with the Extended Bargmann algebra. The Chern--Simons theories based on these algebras can therefore be viewed as spin-3 extensions of Extended Bargmann gravity, studied in \cite{Papageorgiou:2009zc,Papageorgiou:2010ud,Bergshoeff:2016lwr,Hartong:2016yrf}. In the next subsections, we will discuss these spin-3 Extended Bargmann gravity theories, at the linearized level.

\subsubsection{Chern--Simons Theory for $\mathfrak{hs}_3\mathfrak{ebarg1}$}

Using the structure constants of table \ref{tab:ebarg1} and the invariant metric (\ref{eq:trebarg1}), the Chern--Simons action for $\mathfrak{hs}_3\mathfrak{ebarg1}$ can be constructed. As before, we will restrict ourselves  to a linearized analysis. In particular, one can show that the following
\begin{equation}
   \label{eq:flatsol2}
   \bar{A}_\mu = \delta_\mu^0\, \Ht + \delta_\mu^a\, \Pt_a \,,
 \end{equation}
is again a solution of the full non-linear equations of motion. Similar in spirit to (\ref{eq:gaugefieldlin}), we consider the following linearization ansatz
\begin{align}
   \label{eq:gaugefieldlinebarg1}
 A_\mu &= \left(\delta_\mu^0 + \tau_\mu \right) \, \Ht + \left( \delta_\mu^a + e_\mu{}^a \right) \, \Pt_a + \omega_\mu \, \Jt + B_\mu{}^a \, \Gt_a + \tau_\mu{}^a \, \Ht_a + e_\mu{}^{ab} \, \Pt_{ab} + \omega_\mu{}^a \, \Jt_a + B_\mu{}^{ab} \, \Gt_{ab} \nonumber \\ & \quad + \tau_\mu^* \, \Ht^* + \omega_\mu^* \, \Jt^* + \tau_\mu{}^{a *} \, \Ht^*_a + \omega_\mu{}^{a *} \, \Jt_a^* \,.
\end{align}
The linearized field equations can  be obtained by putting the following linearized field strengths to zero
 \begin{align}
   \label{eq:lincovcurvsebarg1}
   R_{\mu\nu}(\Ht) &= \partial_\mu \tau_\nu - \partial_\nu \tau_\mu \,, \nonumber \\
   R_{\mu\nu}(\Pt_a) &= \partial_\mu e_\nu{}^a - \partial_\nu e_\mu{}^a - \epsilon^{ab} \delta_\mu^0\, B_{\nu b} + \epsilon^{ab} \delta_\nu^0\, B_{\mu b} + \epsilon^{ab} \delta_{\mu b}\, \omega_\nu - \epsilon^{ab} \delta_{\nu b}\, \omega_\mu \,, \nonumber \\
   R_{\mu\nu}(\Jt) &= \partial_\mu \omega_\nu - \partial_\nu \omega_\mu \,, \nonumber \\
   R_{\mu\nu}(\Gt^a) &= \partial_\mu B_\nu{}^a - \partial_\nu B_\mu{}^a \,, \nonumber \\
   R_{\mu\nu}(\Ht^a) &= \partial_\mu \tau_\nu{}^a - \partial_\nu \tau_\mu{}^a - \epsilon^{ab} \delta_\mu^0 \omega_{\nu b} + \epsilon^{ab} \delta_\nu^0 \omega_{\mu b} \,, \nonumber \\
   R_{\mu\nu}(\Pt^{ab}) &= \partial_\mu e_\nu{}^{ab} - \partial_\nu e_\mu{}^{ab} - \delta_\mu^c \omega_\nu{}^d \epsilon_{cd} \delta^{ab} + \delta_\nu^c \omega_\mu{}^d \epsilon_{cd} \delta^{ab} - \frac12 \delta_{\mu c} \epsilon^{c(a} \omega_\nu{}^{b)} + \frac12 \delta_{\nu c} \epsilon^{c(a} \omega_\mu{}^{b)} \nonumber \\ & \quad + \delta_\mu^0 \epsilon^{c(a} B_\nu{}^{b)c} - \delta_\nu^0 \epsilon^{c(a} B_\mu{}^{b)c} \,, \nonumber \\
   R_{\mu\nu}(\Jt^a) &= \partial_\mu \omega_\nu{}^a - \partial_\nu \omega_\mu{}^a \,, \nonumber \\
   R_{\mu\nu}(\Gt^{ab}) &= \partial_\mu B_\nu{}^{ab} - \partial_\nu B_\mu{}^{ab} \,, \nonumber \\
   R_{\mu\nu}(\Ht^*) &= \partial_\mu \tau_\nu^* - \partial_\nu \tau_\mu^* \,, \nonumber \\
   R_{\mu\nu}(\Jt^*) &= \partial_\mu \omega_\nu^* - \partial_\nu \omega_\mu^* + \delta_\mu^a B_\nu{}^b \epsilon_{ab} - \delta_\nu^a B_\mu{}^b \epsilon_{ab} \,, \nonumber \\
   R_{\mu\nu}(\Ht^{a *}) &= \partial_\mu \tau_\nu{}^{a *} - \partial_\nu \tau_\mu{}^{a *} \,, \nonumber \\
   R_{\mu\nu}(\Jt^{a *}) &= \partial_\mu \omega_\nu{}^{a *} - \partial_\nu \omega_\mu{}^{a *} +2 \delta_\mu^b B_\nu{}^{ca} \epsilon_{bc} - 2  \delta_\nu^b B_\mu{}^{ca} \epsilon_{bc} - \epsilon^{ab} \delta_\mu^0 \tau_{\nu b}^* + \epsilon^{ab} \delta_\nu^0 \tau_{\mu b}^* \,.
\end{align}
 As in the Carroll cases, the spin connections $\omega_\mu$, $B_\mu{}^a$, $\omega_\mu{}^a$ and $B_\mu{}^{ab}$ appear algebraically in the equations
 \begin{equation}
   R_{\mu\nu}(\Pt^a) = 0 \,, \qquad R_{\mu\nu}(\Jt^*) = 0 \,, \qquad R_{\mu\nu}(\Ht^a) = 0 \,, \qquad R_{\mu\nu}(\Pt^{ab}) = 0 \,, \qquad R_{\mu\nu}(\Jt^{a *}) = 0 \,,
\end{equation}
and one can use these equations to express the spin connections in terms of the other fields. In the spin-2 sector, one straightforwardly finds that $R_{\mu\nu}(\Pt^a) = 0$ and $R_{\mu\nu}(\Jt^*) = 0$ lead to
\begin{align}
   \label{eq:solspin2connebarg1}
   \omega_\mu &= \frac12 \delta_\mu^0 \left( \epsilon^{ab} \left(\partial_a e_0{}^ b - \partial_0 e_a{}^b\right) + \frac12 \epsilon^{ab} \left(\partial_a \omega_b^* - \partial_b \omega_a^*\right) \right) + \frac12 \delta_\mu^a \epsilon^{bc} \left( \partial_b e_a{}^c + \partial_b e_c{}^a - \partial_a e_{b}{}^c \right) \,, \nonumber \\
   B_\mu{}^a &= \delta_\mu^0 \epsilon^{ab} \left( \partial_b \omega_0^* - \partial_0 \omega_b^*\right) + \frac12 \delta_\mu^b \epsilon^{ac} \left(\partial_{(b} e_{|0|c)} - \partial_0 e_{(bc)} - \partial_b \omega_c^* + \partial_c \omega_b^* \right) \,.
 \end{align}
Note that all components of $\omega_\mu$ and $B_\mu{}^a$ are now uniquely determined. This follows from the fact that the system of equations $R_{\mu\nu}(\Pt^a) = 0$ and $R_{\mu\nu}(\Jt^*) = 0$ contains as many independent equations as there are independent components of $\omega_\mu$ and $B_\mu{}^a$.

In the spin-3 sector, one can use the equations
\begin{equation} \label{eq:overdet1}
  R_{\mu\nu}(\Ht^a) = 0 \,, \qquad R_{\mu\nu}(\Pt^{ab}) = 0 \,, \qquad R_{\mu\nu}(\Jt^{a*}) = 0 \,,
\end{equation}
to obtain expressions for the connections $\omega_\mu{}^a$ and $B_\mu{}^{ab}$. Unlike for the spin-2 sector, the system of equations (\ref{eq:overdet1}) is overdetermined, i.e. the number of equations exceeds the number of independent components of $\omega_\mu{}^a$ and $B_\mu{}^{ab}$. One can choose some of the equations of (\ref{eq:overdet1}) to express $\omega_\mu{}^a$ and $B_\mu{}^{ab}$ in terms of $\tau_\mu{}^a$, $e_\mu{}^{ab}$, $\omega_\mu{}^{a *}$ and their derivatives. The remaining equations can then be interpreted as geometrical constraints, once the expressions for the spin connections are used. Dividing the system (\ref{eq:overdet1}) into equations that are solved for connections and equations that remain as constraints, can not be done in a unique way. One convenient way is to use
\begin{equation}
  R_{0b}(\Ht^a) = 0 \,, \qquad R_{0b}(\Pt^{aa}) = 0 \,,
\end{equation}
to solve for $\omega_\mu{}^a$ and
\begin{equation}
  R_{0b}(\Jt^{a *}) - \frac12 \delta_b^a R_{0c}(\Jt^{c *}) = 0 \,, \qquad R_{bc}(\Jt^{a *}) = 0 \,, \qquad R_{0c}(\Pt^{ab}) - \frac12 \delta^{ab} R_{0c}(\Pt^{dd}) = 0 \,,
\end{equation}
to solve for $B_\mu{}^{ab}$, while the remaining equations can be interpreted as geometrical constraints. The solutions for the connections are then explicitly given by
\begin{align}
  \label{eq:spin3connebarg1}
  \omega_\mu{}^a &= \frac13 \delta_\mu^0 \epsilon^{ab} \left(\partial_0 e_b{}^{cc} - \partial_b e_0{}^{cc} \right) + \delta_\mu^b \epsilon^{ac} \left( \partial_b \tau_0{}^c - \partial_0 \tau_b{}^c\right) \,, \nonumber \\
  B_\mu{}^{ab} &= \frac14 \delta_\mu^0 \epsilon^{c(a} \left(\partial_0 \omega_c{}^{b)*} - \partial_c \omega_0{}^{b)*} \right) - \frac14 \delta_\mu^0 \tau^{(ab)*} + \frac12 \delta_\mu^0 \delta^{ab} \tau_c{}^{c*} \nonumber \\ & \quad + \frac14 \delta_\mu^c \epsilon^{d(a} \left( \partial_0 e_d{}^{b)c} - \partial_d e_0{}^{b)c} \right) - \frac14 \delta_\mu^c \delta_c^{(a} \epsilon^{|de|} \left(\partial_0 e_d{}^{b)e} - \partial_d e_0{}^{b)e} \right) \nonumber \\ & \quad + \frac34 \delta_\mu^c \delta^{ab} \epsilon^{cd} \left(\partial_0 e_d{}^{ee} - \partial_d e_0{}^{ee} \right) - \frac12 \delta_\mu^c \delta^{ab} \epsilon^{ed} \left(\partial_0 e_d{}^{ec} - \partial_d e_0{}^{ec} \right) \nonumber \\ & \quad - \frac14 \delta_\mu^c \delta^{ab} \epsilon^{de} \left( \partial_d \omega_e{}^{c *} - \partial_e \omega_d{}^{c *} \right) - \frac16 \delta_\mu^c \delta^{c(a} \epsilon^{b)d} \left(\partial_0 e_d{}^{ee} - \partial_d e_0{}^{ee} \right) \,. 
\end{align}

\subsubsection{Chern--Simons Theory for $\mathfrak{hs}_3\mathfrak{ebarg2}$}

The Chern--Simons theory for $\mathfrak{hs}_3\mathfrak{ebarg2}$, with structure constants given in table \ref{tab:ebarg1} and invariant metric (\ref{eq:trebarg2}), can again be linearized using the linearization ansatz (\ref{eq:gaugefieldlinebarg1}). The resulting linearized equations of motion are now given by
\begin{align}
  \label{eq:lineomsebarg2}
 0 &=  R_{\mu\nu}(\Ht) \equiv \partial_\mu \tau_\nu - \partial_\nu \tau_\mu \,, \nonumber \\
 0 &=  R_{\mu\nu}(\Pt^a) \equiv \partial_\mu e_\nu{}^a - \partial_\nu e_\mu{}^a - \epsilon^{ab} \delta_\mu^0 B_{\nu b} + \epsilon^{ab} \delta_\nu^0 B_{\mu b} + \epsilon^{ab} \delta_{\mu b} \omega_\nu - \epsilon^{ab} \delta_{\nu b} \omega_\mu \,, \nonumber \\
 0 &=  R_{\mu\nu}(\Jt) \equiv \partial_\mu \omega_\nu - \partial_\nu \omega_\mu \,, \nonumber \\
 0 &=  R_{\mu\nu}(\Gt^a) \equiv \partial_\mu B_\nu{}^a - \partial_\nu B_\mu{}^a \,, \nonumber \\
 0 &=  R_{\mu\nu}(\Ht^a) \equiv \partial_\mu \tau_\nu{}^a - \partial_\nu \tau_\mu{}^a - \epsilon^{ab} \delta_\mu^0 \omega_{\nu b} + \epsilon^{ab} \delta_\nu^0 \omega_{\mu b} - 2 \delta_\mu^b B_\nu{}^{ca} \epsilon_{bc} + 2 \delta_\nu^b B_\mu{}^{ca} \epsilon_{bc} \,, \nonumber \\
 0 &=  R_{\mu\nu}(\Pt^{ab}) \equiv \partial_\mu e_\nu{}^{ab} - \partial_\nu e_\mu{}^{ab} + \delta_\mu^0 \epsilon^{c(a} B_\nu{}^{b)c} - \delta_\nu^0 \epsilon^{c(a} B_\mu{}^{b)c} \,, \nonumber \\
 0 &=  R_{\mu\nu}(\Jt^a) \equiv \partial_\mu \omega_\nu{}^a - \partial_\nu \omega_\mu{}^a \,, \nonumber \\
 0 &=  R_{\mu\nu}(\Gt^{ab}) \equiv \partial_\mu B_\nu{}^{ab} - \partial_\nu B_\mu{}^{ab} \,, \nonumber \\
 0 &=  R_{\mu\nu}(\Ht^*) \equiv \partial_\mu \tau_\nu^* - \partial_\nu \tau_\mu^* \,, \nonumber \\
 0 &=  R_{\mu\nu}(\Jt^*) \equiv \partial_\mu \omega_\nu^* - \partial_\nu \omega_\mu^* + \delta_\mu^a B_\nu{}^b \epsilon_{ab} - \delta_\nu^a B_\mu{}^b \epsilon_{ab} \,, \nonumber \\
 0 &=  R_{\mu\nu}(\Pt^{ab *}) \equiv \partial_\mu e_\nu{}^{ab *} - \partial_\nu e_\mu{}^{ab *} \,, \nonumber \\
 0 &=  R_{\mu\nu}(\Gt^{ab *}) \equiv \partial_\mu B_\nu{}^{ab *} - \partial_\nu B_\mu{}^{ab *} + \delta_\mu^0 \epsilon^{c(a} e_\nu{}^{b)c *} - \delta_\nu^0 \epsilon^{c(a} e_\mu{}^{b)c *}\nonumber \\ & \qquad \qquad \qquad \ \  - \frac12 \epsilon^{c(a} \delta_{\mu c} \omega_\nu{}^{b)} + \frac12 \epsilon^{c(a} \delta_{\nu c} \omega_\mu{}^{b)} \,.
\end{align}
One can now use the equations
\begin{equation}
  R_{\mu\nu}(\Pt^a) = 0 \,, \quad R_{\mu\nu}(\Jt^*) = 0 \,, \quad R_{\mu\nu}(\Ht^a) = 0\,, \quad R_{\mu\nu}(\Pt^{ab}) = 0 \,, \quad R_{\mu\nu}(\Gt^{ab *}) = 0 \,,
\end{equation}
to express the connections $\omega_\mu$, $B_\mu{}^a$, $\omega_\mu{}^a$ and $B_\mu{}^{ab}$ in terms of the other fields. For the spin-2 sector fields $\omega_\mu$, $B_\mu{}^a$, this is done by solving
\begin{equation}
  R_{\mu\nu}(\Pt^a) = 0 \,, \qquad R_{\mu\nu}(\Jt^*) = 0 \,.
\end{equation}
These equations are the same as in the case of $\mathfrak{hs}_3\mathfrak{ebarg1}$ and one finds the same solution (\ref{eq:solspin2connebarg1}). For the spin-3 sector, one has to solve the equations
\begin{equation}
  R_{\mu\nu}(\Ht^a) = 0\,, \qquad R_{\mu\nu}(\Pt^{ab}) = 0 \,, \qquad R_{\mu\nu}(\Gt^{ab *}) = 0 \,.
\end{equation}
As in the $\mathfrak{hs}_3\mathfrak{ebarg1}$ case, this is an overdetermined set of equations, that can be used to obtain expressions for $\omega_\mu{}^a$ and $B_\mu{}^{ab}$ along with geometrical constraints. One convenient way of doing this, is by using the equations
\begin{equation}
  R_{0a}(\Ht^a) = 0 \,, \qquad \epsilon^{bc} R_{0c}(\Gt^{ba *}) = 0 \,, \qquad R_{cd}(\Gt^{ab *}) = 0 \,,
\end{equation}
to solve for $\omega_\mu{}^a$. An expression for $B_\mu{}^{ab}$ can be found as a solution of
\begin{equation}
  R_{0b}(\Ht^a) - \frac12 \delta_b^a R_{0c}(\Ht^c) = 0 \,, \qquad R_{bc}(\Ht^a) = 0 \,, \qquad R_{0c}(\Pt^{ab}) - \frac12 \delta^{ab} R_{0c}(\Pt^{dd}) = 0 \,.
\end{equation}
The connections $\omega_\mu{}^a$ and $B_\mu{}^{ab}$ are then given by
\begin{align}
  \label{eq:spin3connebarg2}
  \omega_\mu{}^a &= -\frac23 \delta^0_\mu \left(\epsilon^{bc} \left(\partial_0 B_b{}^{ca *} - \partial_b B_0{}^{ca *}\right) + 2 e_b{}^{ab *} - e_a{}^{bb *} \right) + \frac12 \delta_\mu^b \epsilon^{cd} \left(\partial_c B_d{}^{ab*} - \partial_d B_c{}^{ab*} \right) \nonumber \\ & \quad \, - \frac12 \delta_\mu^b \epsilon^{ab} \left(\partial_0 \tau_c{}^c - \partial_c \tau_0{}^c \right) \,, \nonumber \\
  B_\mu{}^{ab} &= \frac14 \delta_\mu^0 \epsilon^{(a|c|} \left(\partial_0 \tau_c{}^{b)} - \partial_c \tau_0{}^{b)}\right) + \frac14 \delta_\mu^0 \epsilon^{cd} \left(\partial_c B_d{}^{ab*} - \partial_d B_c{}^{ab*} \right) \nonumber \\ & \quad \, - \frac14 \delta_\mu^0 \delta^{ab} \epsilon^{cd} \left(\partial_c B_d{}^{ee*} - \partial_d B_c{}^{ee*} \right) + \frac14 \delta_\mu^c \epsilon^{d(a} \left(\partial_0 e_c{}^{b)d} - \partial_c e_0{}^{b)d}  \right) \nonumber \\ & \quad \, + \frac12 \delta_\mu^c \delta^{ab} \epsilon^{de} \left(\partial_0 e_d{}^{ce} - \partial_d e_0{}^{ce}\right) + \frac14 \delta_\mu^c \delta^{ab} \epsilon^{cd} \left(\partial_0 e_d{}^{ee} - \partial_d e_0{}^{ee}\right) \nonumber \\ & \quad \, + \frac14 \delta_\mu^c \delta^{ab} \epsilon^{de} \left( \partial_d \tau_e{}^c - \partial_e \tau_d{}^c\right) \,.
\end{align}

\section{Carroll Gravity as Example}\label{sec:4}

\newcommand{\boost}{B}

In this section we address whether there are interesting infinite extensions of the algebras discussed above, in the same way that the global conformal algebra in two dimensions gets extended to the Virasoro algebra by imposing Brown--Henneaux boundary conditions \cite{Brown:1986nw}. Rather than being as comprehensive as in the other sections we focus here on a specific simple example. In fact, we drop the higher spin fields and consider spin-2 Carroll gravity, defined by a Chern--Simons gauge theory with action \eqref{eq:CSaction} where the connection 1-form
\eq{
A =  \tau\, \Ht + e^a\, \Pt_a  + \omega\, \Jt + \boost^a\, \Gt_a
}{eq:car1}
takes values in the spin-2 Carroll algebra ($a=1,2$), whose non-vanishing commutation relations read
\begin{subequations}
\label{eq:car2}
\begin{align}
    [\Jt,\,\Pt_a] &= \epsilon_{ab}\, \Pt_b\,, \\
    [\Jt,\,\Gt_a] &= \epsilon_{ab}\, \Gt_b\,, \\
    [\Pt_a,\,\Gt_b] &= -\epsilon_{ab}\, \Ht\,,
\end{align}
\end{subequations}
where we use the convention $\epsilon_{12}=+1$ for the antisymmetric $\epsilon$-symbol. The invariant metric has the non-vanishing entries
\eq{
\langle \Ht,\Jt\rangle = -1 \qquad \langle \Pt_a,\Gt_b\rangle = \delta_{ab}
}{eq:car3}
fully compatible with \eqref{eq:invmetrichscar1} in the absence of higher spin generators.

Our main goal is not just to find some infinite extension of the algebra \eqref{eq:car2} (this always exists at least in the form of the loop algebra of the underlying gauge algebra, see e.g.~\cite{Elitzur:1989nr}; for AdS$_3$ gravity such boundary conditions were investigated recently in \cite{Grumiller:2016pqb}), but rather to find an extension that has a `nice' geometric interpretation along the lines of the Brown--Henneaux boundary conditions. This means that we want to achieve a suitable Drinfeld--Sokolov type of reduction where not all algebraic components of the connection are allowed to fluctuate. The words `nice' and `suitable' here mean that,
in particular,  we want that  the appropriate Carroll background geometry as part of our spectrum of physical states is allowed by our boundary conditions, and that all additional states are fluctuations around this background. First, we recall  some basic aspects of Carroll geometry.

The Carroll-zweibein for the flat background geometry in some Fefferman--Graham like coordinates should take the form
\eq{
e^1_\vp = \rho\qquad e^2_\rho = 1\qquad e^1_\rho=e^2_\vp = 0
}{eq:car4}
so that the corresponding two-dimensional line-element reads
\eq{
\extd s^2_{(2)} = e^a e^b \delta_{ab} = \rho^2\extd\vp^2 + \extd\rho^2\,.
}{eq:car5}
We shall refer to $\rho$ as `radial coordinate' and to $\vp$ as `angular coordinate', assuming $\vp\sim\vp+2\pi$. Moreover, on the background the time-component should be fixed as
\eq{
\tau = \extd t\,.
}{eq:car6}
Below we shall allow subleading (in $\rho$) fluctuations in the two-dimensional line-element \eqref{eq:car5} and leading fluctuations in the time-component \eqref{eq:car6}.

We proceed now by stating the result for the boundary conditions that define our example of Carroll gravity and discuss afterwards the rationale behind our choices as well as the consistency of the boundary conditions by proving the finiteness, integrability, non-triviality and conservation of the canonical boundary charges. We follow the general recipe reviewed e.g.~in \cite{Afshar:2012nk, Riegler:2016hah}. First, we bring the connection \eqref{eq:car1} into a convenient gauge (see for instance \cite{Banados:1998gg})
\eq{
A = b^{-1}(\rho)\,\big(\extd+a(t,\,\vp)\big)\,b(\rho)
}{eq:car7}
where the group element
\eq{
b(\rho) = e^{\rho \Pt_2}
}{eq:car8}
is fixed as part of the specification of our boundary conditions, $\delta b=0$. The boundary connection $a$ does not depend on the radial coordinate $\rho$ and is given by
\begin{subequations}
\label{eq:car9}
\begin{align}
a_\vp &= -\Jt + h(t,\,\vp)\, \Ht + p_a(t,\,\vp)\, \Pt_a + g_a(t,\,\vp)\, \Gt_a\,,  \label{eq:car9a} \\
a_t &= \mu(t,\,\vp)\, \Ht\,, \label{eq:car9b}
\end{align}
\end{subequations}
where $\mu$ is arbitrary but fixed, $\delta\mu=0$, while all other functions are arbitrary and can vary. This means that the allowed variations of the boundary connection are given by
\eq{
\delta a = \delta a_\vp\,\extd\vp = \big(\delta h\, \Ht + \delta p_a \,\Pt_a + \delta g_a\, \Gt_a \big)\,\extd\vp\,.
}{eq:car10}
The full connection in terms of the boundary connection is then given by
\eq{
A = a + \Pt_2 \extd\rho + \rho\, [a,\,\Pt_2]
}{eq:car7too}
and acquires its non-trivial radial dependence through the last term, $\rho\, [a,\,\Pt_2]=\rho\,(\Pt_1 - g_1(t,\,\vp)\,\Ht)\,\extd\vp$. Only the $\vp$-component of the connection is then allowed to vary.
\eq{
\delta A = \delta a + \rho\, [\delta a,\,\Pt_2] = \big(\delta h\, \Ht  + \delta p_a\, \Pt_a + \delta g_a\, \Gt_a - \rho\,\delta g_1\, \Ht \big)\,\extd\vp
}{eq:car10too}

The above boundary conditions lead to Carroll-geometries of the form
\eq{
\extd s^2_{(2)} = \big[\big(\rho + p_1(t,\,\vp)\big)^2 + p_2(t,\,\vp)^2\big]\,\extd\vp^2 + 2p_2(t,\,\vp)\,\extd\vp\extd\rho + \extd\rho^2
}{eq:car11}
and
\eq{
\tau = \mu(t,\,\vp)\,\extd t + \big(h(t,\,\vp) - \rho\, g_1(t,\,\vp)\big)\,\extd\vp\,.
}{eq:car12}
Thus, we see that to leading order in $\rho$ the background line-element \eqref{eq:car5} is recovered from \eqref{eq:car11}, plus subleading (state-dependent) fluctuations captured by the functions $p_a(t,\,\vp)$. As we shall see in the next paragraph the functions $p_a$ and $g_a$ are $t$-independent on-shell. In the metric-formulation our boundary conditions can be phrased as
\eq{
\extd s^2_{(2)} = \big(\rho^2 + {\cal O}(\rho)\big)\,\extd\vp^2 + {\cal O}(1)\,\extd\rho\extd\vp + \extd\rho^2
}{eq:car27}
and
\eq{
\tau = \mu(t,\,\vp)\,\extd t + {\cal O}(\rho)\,\extd\vp\,.
}{eq:car29}
Note that while the asymptotic form of the two-dimensional line-element \eqref{eq:car27} may have been guessed easily, the specific form of the time-component \eqref{eq:car29} is much harder to guess, particularly the existence of a `shift'-component proportional to $\extd\vp$ that grows linearly in $\rho$. Fortunately, the Chern--Simons formulation together with the gauge choice \eqref{eq:car7} minimizes the amount of guesswork needed to come up with meaningful boundary conditions.

We consider now the impact of the equations of motion on the free functions in the boundary connection \eqref{eq:car9}. Gauge-flatness $F=0$ implies
\eq{
\partial_t a_\vp - \partial_\vp a_t + [a_t,\,a_\vp]=\partial_t a_\vp - \partial_\vp a_t = 0\,.
}{eq:car13}
As a consequence, we get the on-shell conditions (which also could be called `holographic Ward identities')
\eq{
\partial_t p_a = \partial_t g_a = 0\qquad \partial_t h = \partial_\vp\mu\,.
}{eq:car14}
Thus, most of the functions in the boundary connection \eqref{eq:car9} are time-independent, with the possible exception of $h$ and $\mu$.

The boundary-condition preserving transformations, $\delta_{\hat\lambda} A = \extd\hat\lambda + [A,\,\hat\lambda] = {\cal O}(\delta A)$, generated by $\hat\lambda=b^{-1}\lambda b$ have to obey the relations
\begin{subequations}
\label{eq:car15}
\begin{align}
  \delta_\lambda a_t &= \partial_t\lambda + [a_t,\,\lambda] = \partial_t \lambda = 0\,,\ \label{eq:car15a} \\
  \delta_\lambda a_\vp &= \partial_t\lambda + [a_\vp,\,\lambda] = {\cal O}(\delta a_\vp)\,, \label{eq:car15b}
\end{align}
\end{subequations}
where $ {\cal O}(\delta a_\vp)$ denotes all the allowed variations displayed in \eqref{eq:car10}. It is useful to decompose $\lambda$ with respect to the algebra \eqref{eq:car1}.
\eq{
\lambda = \lambda^\Ht\, \Ht + \lambda^{\Pt_a}\, \Pt_a + \lambda^\Jt\,\Jt + \lambda^{\Gt_a}\, \Gt_a\,.
}{eq:car18}
The first line in \eqref{eq:car15} establishes the time-independence of $\lambda$, while the second line yields the consistency condition
\eq{
\partial_\vp \lambda^\Jt = 0
}{eq:car16}
as well as the transformations rules
\begin{subequations}
\label{eq:car17}
\begin{align}
 \delta_\lambda h &= \partial_\vp \lambda^\Ht - \big(p_1 \lambda^{\Gt_2} - p_2 \lambda^{\Gt_1} + g_1 \lambda^{\Pt_2} -  g_2 \lambda^{\Pt_1}\big)\,,\\
 \delta_\lambda p_a &= \partial_\vp \lambda^{\Pt_a} - \eps_{ab} \big(\lambda^{\Pt_b} - p_b\lambda^\Jt\big)\,,\\
 \delta_\lambda g_a &= \partial_\vp \lambda^{\Gt_a} - \eps_{ab} \big(\lambda^{\Gt_b} - g_b\lambda^\Jt\big)\,.
\end{align}
\end{subequations}

Applying the Regge--Teitelboim approach \cite{Regge:1974zd} to Chern--Simons theories yields the following background-independent result for the variation of the canonical boundary charges
\eq{
\delta Q[\lambda] = \frac{k}{2\pi}\,\oint\langle\hat\lambda\, \delta A\rangle = \frac{k}{2\pi}\,\oint\langle\lambda\, \delta a_\vp\rangle\,\extd\vp
}{eq:car19}
which in our case expands to
\eq{
\delta Q[\lambda] = \frac{k}{2\pi}\,\oint\big(-\lambda^\Jt \delta h + \lambda^{\Pt_a} \delta g_a + \lambda^{\Gt_a} \delta p_a\big)\,\extd\vp\,.
}{eq:car20}
The canonical boundary charges are manifestly finite since the $\rho$-dependence drops out in \eqref{eq:car19}; they are also integrable in field-space since our $\lambda$ is state-independent.
\eq{
Q[\lambda] = \frac{k}{2\pi}\,\oint\big(-\lambda^\Jt h + \lambda^{\Pt_a} g_a + \lambda^{\Gt_a} p_a\big)\,\extd\vp\,.
}{eq:car21}
The result \eqref{eq:car21} clearly is non-trivial in general. To conclude the proof that we have meaningful boundary conditions we finally check conservation in time, using the on-shell relations \eqref{eq:car14} as well as the time-independence of $\lambda$, see \eqref{eq:car15a}:
\eq{
\partial_t Q[\lambda]\big|_{\textrm{\tiny EOM}} = -\frac{k}{2\pi}\,\oint \lambda^\Jt \partial_t h\,\extd\vp = -\frac{k}{2\pi}\,\oint \lambda^\Jt \partial_\vp \mu\,\extd\vp = \frac{k}{2\pi}\,\oint \mu \partial_\vp \lambda^\Jt \,\extd\vp\,.
}{eq:car22}
By virtue of \eqref{eq:car16} we see that the last integrand vanishes and thus we have established charge conservation on-shell:
\eq{
\partial_t Q[\lambda]\big|_{\textrm{\tiny EOM}}=0\,.
}{eq:car23}
Since our canonical boundary charges \eqref{eq:car21} are finite, integrable in field space, non-trivial and conserved in time the boundary conditions \eqref{eq:car7}-\eqref{eq:car10too} are consistent and lead to a non-trivial theory. For later purposes, it is useful to note that due to the constancy of $\lambda^\Jt$ only the zero mode charge associated with the function $h$ can be non-trivial. This means that we can gauge-fix our connection using proper gauge transformations such that $h=\rm const.$

 We now introduce Fourier modes in order to be able to present the asymptotic symmetry algebra in a convenient form.\footnote{%
There is no meaning to the index positions in this section. The only reason why we write $\Pt^a_n$ and $\Gt^a_n$ instead of corresponding quantities with lower indices is that our current convention is easier to read.
}
\begin{subequations}
\label{eq:car24}
\begin{align}
  \Pt^a_n &:= \frac{1}{2\pi}\,\oint \extd\vp\, e^{in\vp}g_a(t,\,\vp)\big|_{\textrm{\tiny EOM}}\,, \\
  \Gt^a_n &:= \frac{1}{2\pi}\,\oint \extd\vp\, e^{in\vp}p_a(t,\,\vp)\big|_{\textrm{\tiny EOM}}\,, \\
  \Jt &:= -\frac{1}{2\pi}\,\oint \extd\vp\, h(t,\,\vp)\big|_{\textrm{\tiny EOM}}\,.
\end{align}
\end{subequations}
A few explanations are in order. Due to our off-diagonal bilinear form \eqref{eq:car3} we associate the $n^{\textrm{th}}$ Fourier mode of the functions $g_a$ ($p_a$) with the generator $\Pt^a_n$ ($\Gt^a_n$). For the same reason we associate $\Jt$ with minus the zero mode of $h$. Finally, the subscript `EOM' means that all integrals are evaluated on-shell, in which case all $t$-dependence drops out (and in the last integral also all $\vp$-dependence).

We make a similar Fourier decomposition of the gauge parameters $\lambda^i$, where $i$ refers to the generators $\Pt_a$, $\Gt_a$ and $\Jt$; the parameter $\lambda^\Ht$ is not needed since it does not appear in the canonical boundary charges \eqref{eq:car21}, so all gauge transformations associated with it are proper ones and can be used to make $h$ constant.
\begin{subequations}
\label{eq:car25}
\begin{align}
  \lambda^{\Pt_a}_n &:= \frac{1}{2\pi}\,\oint \extd\vp\, e^{in\vp}\lambda^{\Pt_a}(\vp)\,, \\
  \lambda^{\Gt_a}_n &:= \frac{1}{2\pi}\,\oint \extd\vp\, e^{in\vp}\lambda^{\Gt_a}(\vp)\,.
\end{align}
\end{subequations}
Note that we have used \eqref{eq:car15} to eliminate all time-dependence and that $\lambda^\Jt$ is a constant according to \eqref{eq:car16} thus requiring no Fourier decomposition.

The variations \eqref{eq:car17} of the state-dependent functions then establish corresponding variations in terms of the Fourier components \eqref{eq:car24}, \eqref{eq:car25}.
\begin{subequations}
\label{eq:car26}
\begin{align}
  \delta \Pt_n^a &= -in \lambda^{\Gt_a}_n - \eps_{ab} \lambda^{\Gt_b}_n + \eps_{ab}\lambda^\Jt \Pt^b_n\,,\\
  \delta \Gt_n^a &= -in \lambda^{\Pt_a}_n - \eps_{ab} \lambda^{\Pt_b}_n + \eps_{ab}\lambda^\Jt \Gt^b_n\,,\\
  \delta \Jt &= \sum_{n\in\mathbb{Z}} \eps_{ab}\Big(\Gt^a_n\lambda^{\Gt_b}_{-n} + \Pt^a_n \lambda_{-n}^{\Pt_b}\Big)\,.
\end{align}
\end{subequations}
From the variations \eqref{eq:car26} we can read off the asymptotic symmetry algebra, using the  fact that the canonical generators generate gauge transformations via the Dirac bracket $\delta_{\lambda_1} Q[\lambda_2]=\{Q[\lambda_1],\,Q[\lambda_2]\}$.

Converting Dirac brackets into commutators then establishes the asymptotic symmetry algebra as the commutator algebra of the infinite set of generators $\Pt^a_n$, $\Gt^a_n$ and $\Jt$. The central element of this algebra will be associated with (minus) $\Ht$, concurrent with the notation of \eqref{eq:car2}. Evaluating \eqref{eq:car26} yields\footnote{%
Note that our definitions of Fourier-components \eqref{eq:car24}, \eqref{eq:car25} require that we associate the negative Fourier components of the $\lambda$ with the positive Fourier components of the generators so that, for instance, $[\Pt^b_n,\,\Jt]=\delta_{\lambda^{\Pt_b}_{-n}} \Jt$.
}
\begin{subequations}
\label{eq:car28}
\begin{align}
  [\Jt,\,\Pt^a_n] &= \eps_{ab}\,\Pt^b_n\,, \\
  [\Jt,\,\Gt^a_n] &= \eps_{ab}\,\Gt^b_n\,, \\
  [\Pt^a_n,\,\Gt^b_m] &= -\big(\eps_{ab} + in \delta_{ab}\big)\,\Ht\,\delta_{n+m,\,0}\,,
\end{align}
\end{subequations}
where all commutators not displayed vanish. We have thus succeeded in providing an infinite lift of the Carroll algebra \eqref{eq:car2}, which is contained as a subalgebra of our asymptotic symmetry algebra \eqref{eq:car28} by restricting to the zero-mode generators $\Pt_a = \Pt^a_0$, $\Gt_a= \Gt^a_0$ in addition to $\Jt$ and $\Ht$. As a simple consistency check one may verify that the Jacobi identities indeed hold. The only non-trivial one to be checked is the identity $[[\Jt,\,\Pt^a_n],\,\Gt_m^b]+\textrm{cycl.} = 0$.

We conclude this section with a couple of remarks. The boundary conditions \eqref{eq:car7}-\eqref{eq:car9} by no means are unique and can be either generalized or specialized to looser or stricter ones, respectively. In particular, we have switched off nearly all `chemical potentials' in our specification of the time-component of the connection \eqref{eq:car9b}, and it could be of interest to allow arbitrary chemical potentials. Apart from this issue there is only one substantial generalization of our boundary conditions, namely to allow for a state-dependent function in front of the generator $\Jt$ in the angular component of the connection \eqref{eq:car9a}. As mentioned in the opening paragraph of this section, in that case the expected asymptotic symmetry algebra is the loop algebra of the Carroll algebra \eqref{eq:car2}. In principle, it is possible to make our boundary conditions stricter, but that would potentially eliminate interesting physical states like some of the Carroll geometries \eqref{eq:car11}, \eqref{eq:car12}.\footnote{%
Perhaps the most interesting specialization of our boundary conditions would be one where the two-dimensional line-element \eqref{eq:car11} is unaltered but the time-component \eqref{eq:car12} is more restricted, e.g.~by requiring that no shift term proportional to $\extd\vp$ is generated.
} Thus, while our choice \eqref{eq:car7}-\eqref{eq:car9} is not unique it may provide the most interesting set of boundary conditions for spin-2 Carroll gravity. Using the same techniques it should be straightforward to extend the discussion of this section to higher spin Carroll gravity and related theories discussed in this paper.

\section{Discussion}
\label{sec:discussion}

In this paper, we have extended the work of Bacry and L\'evy-Leblond \cite{Bacry:1968zf} by classifying all possible kinematical algebras of three-dimensional theories of a spin-3 field coupled to gravity, that can be obtained via (sequential) In\"on\"u-Wigner contraction procedures of the algebras of spin-3 gravity in (A)dS. This classification can be found in section \ref{sec:algebras} and the resulting possible kinematical algebras, along with their origin via contraction, are  summarized in figure  \ref{fig:hscube}. We have summarized the commutation relations of the algebras in tables \ref{tab:adspoin}-\ref{tab:hsstat2}. The algebras of tables \ref{tab:hscar} and \ref{tab:hsgal} are suitable generalizations of the Carroll and Galilei algebras, that correspond to the ultra-relativistic and non-relativistic limits of the Poincar\'e algebra. These algebras have been used in section \ref{sec:spin-3} as a starting point to construct higher spin generalizations of ultra-relativistic and non-relativistic gravity theories. We have argued that one can easily construct a Chern-Simons action for the spin-3 Carroll algebras, that leads to a spin-3 generalization of Carroll gravity. We have moreover shown that Chern-Simons actions can be written down for suitable extensions of the spin-3 Galilei algebras, that lead to spin-3 generalizations of Extended Bargmann gravity. The algebras constructed in this paper are finite-dimensional. We have shown in section \ref{sec:4} that the three-dimensional  Carroll algebra admits an infinite-dimensional extension, that is the asymptotic symmetry algebra of Carroll gravity with suitable boundary conditions. This can be taken as a hint that similar results hold for the non- and ultra-relativistic algebras constructed in this paper, as well as for the spin-2 algebras whose infinite-dimensional extensions have not been addressed in the literature yet.

There are several questions that are worthwhile for future study. The non- and ultra-relativistic spin-3 gravity theories constructed here, are given in the Chern-Simons (i.e.~first order `zuvielbein') formulation. It is interesting to see whether a metric-like formulation can be constructed and whether the linearized field equations can be rewritten as Fronsdal-like equations. The results for the linearized spin connections given in section \ref{sec:spin-3} should be useful in this regard.

In this paper, we have restricted ourselves to spin-3 theories, by considering algebras that are obtained via (sequential) In\"on\"u-Wigner contraction procedures of $\mathfrak{sl}(3,\mathbb{R}) \oplus \mathfrak{sl}(3,\mathbb{R})$ or $\mathfrak{sl}(3,\mathbb{C})$. This analysis can be extended to theories with fields up to spin $N$, by considering contraction procedures of $\mathfrak{sl}(N,\mathbb{R}) \oplus \mathfrak{sl}(N,\mathbb{R})$ or $\mathfrak{sl}(N,\mathbb{C})$~\cite{Campoleoni:2011hg}. One can then study the non- and ultra-relativistic gravity theories that arise in this way and in particular investigate the types of boundary conditions that lead to interesting asymptotic symmetry algebras. It would be particularly interesting to see whether it is possible to construct non- and ultra-relativistic versions of non-linear $W$-algebras.

Another research direction concerns the inclusion of fermionic fields with spins higher than or equal to 3/2. This will require a classification of contraction procedures of Lie superalgebras and can lead to higher spin generalizations of three-dimensional Extended Bargmann supergravity \cite{Bergshoeff:2016lwr}.

Some of the results presented in this paper are also useful for studies of Ho\v{r}ava-Lifshitz gravity, that has been proposed as a new framework for Lifshitz holography \cite{Griffin:2011xs,Griffin:2012qx,Kiritsis:2012ta,Janiszewski:2012nf,Janiszewski:2012nb,Wu:2014dha,Hartong:2015zia,Hartong:2016yrf}. Extended Bargmann gravity has been argued to correspond to a special case of Ho\v{r}ava-Lifshitz gravity \cite{Hartong:2016yrf}. In this paper, we have constructed spin-3 generalizations of Extended Bargmann gravity. It is conceivable that these can be interpreted as suitable spin-3 generalizations of Ho\v{r}ava-Lifshitz gravity. It would be interesting to check whether this is indeed the case and whether the construction presented here can be generalized to yield spin-3 generalizations of generic Ho\v{r}ava-Lifshitz gravity theories.

Finally, higher spin theory has recently been argued to describe some of the excitations in fractional quantum Hall liquids \cite{Golkar:2016thq}. Newton-Cartan geometry and gravity, that are based on extensions of the Galilei algebra, have been very useful in constructing effective actions that can capture transport properties in studies of the fractional quantum Hall effect. It would be interesting to investigate whether the non-relativistic higher spin gravity theories that can be constructed using the results of this paper, can play a similar role.

\acknowledgments
We thank Matthias Blau, Stefan Fredenhagen, Mirah Gary, Wout Merbis, Radoslav Rashkov, Jakob Salzer, Friedrich Sch\"oller and Massimo Taronna for useful comments.

EB wants to thank the TU Wien for hospitality where part of this work was done. JR and SP want to thank the University of Groningen for hospitality. EB and JR wish to thank the Galileo Galilei Institute for Theoretical Physics for hospitality and the INFN for partial support during the completion of this work.
DG was supported by the Austrian Science Fund (FWF), projects P 27182-N27 and P 28751-N27.
SP was supported by the FWF project P 27396-N27 and also wants to thank the Groningen Lunch Seminar group for interesting questions and comments.
JR was supported by the NCCR SwissMAP, funded by the Swiss National Science Foundation.

\appendix

\section{Conventions}
\label{app:conventions}

In this paper, we adopt the convention that the symmetrization of a pair of indices $a$, $b$ are denoted with parentheses $(a b)$, while anti-symmetrization is denoted with square brackets $[a b]$. Symmetrization and anti-symmetrization is performed without normalization factor, i.e.,
\begin{align}
  T_{(ab)}= T_{ab}+T_{ba} \qquad T_{[ab]}= T_{ab}-T_{ba} \,.
\end{align}
Nested (anti-)symmetrizations are understood to be taken from the outermost ones to the innermost ones, e.g.
\begin{equation}
  T_{(a(bc)d)}= T_{a(bc)d}+ T_{d(bc)a}= T_{abcd}+T_{acbd}+ T_{dbca}+ T_{dcba} \,.
\end{equation}
Vertical bars denote that the (anti-)symmetrization does not affect the enclosed indices, e.g.,
\begin{equation}
  T_{[a|bc|d]}= T_{abcd}-T_{dbca} \,.
\end{equation}
With our conventions this means that $T_{(a|(bc)|d)}= T_{(a(bc)d)}$.

Upper case Latin indices denote space-time indices, while lower case ones denote spatial indices:
\begin{align}
A,B,C,M,\dots&=0,1,2\,, & a,b,c,m,\dots&=1,2 \,.
\end{align}
We take the following conventions for the metric
\begin{align}
  \eta_{AB}=\mathrm{diag}(-,+,+) \qquad  \eta_{ab}=\delta_{ab}=\mathrm{diag}(+,+)\,.
\end{align}
For the Levi-Civita symbol, we adopt the following convention:
 \begin{align}
 \epsilon_{012}=\epsilon_{12}=1\,, \qquad \epsilon_{0ab}=\epsilon_{ab}\,, \qquad \epsilon^{ab} = \epsilon_{ab} \,.
 \end{align}

The Lie algebraic direct sum of the Lie algebras $\mathfrak{a}$ and $\mathfrak{b}$ is denoted by $\mathfrak{a} \oplus \mathfrak{b}$, i.e.~$\mathfrak{a}$ and $\mathfrak{b}$ are ideals. The direct sum as vectorspaces is denoted by $\mathfrak{a} + \mathfrak{b}$, i.e.~they do not necessarily commute with each other.

\section{Proof of Theorem 1}
\label{app:proof}

In order to obtain contraction procedures of $\mathfrak{hs}_{3}\mathfrak{AdS}_{3}$ and $\mathfrak{hs}_{3}\mathfrak{dS}_{3}$ that reduce to those of table \ref{tab:spin2contr} when restricted to the spin-2 part, we start from the subalgebras $\mathfrak{h}$ of table \ref{tab:spin2contr}. For each of these four contraction procedures, one needs to add spin-3 generators to the subalgebra $\mathfrak{h}$, since otherwise one is led to contractions with an abelian spin-3 part. From table \ref{tab:spin2contr} one sees that the subalgebra $\mathfrak{h}$ always contains the generator $\Jt$. The spin-3 generators that one adds to $\mathfrak{h}$ therefore need to fall into irreducible representations of $\Jt$, under the adjoint action, in order to make sure that the enlarged $\mathfrak{h}$ is a subalgebra. The spin-3 generators fall into the following irreducible representations of $\Jt$:
\begin{equation}
  \label{eq:irrepsJ}
  \{\Jt_a\}\,, \quad \{\Ht_a\} \,, \quad \{\Gt_{12},\Gt_{22}-\Gt_{11}\} \,, \quad \{\Gt_{11} + \Gt_{22}\}\,, \quad \{\Pt_{12},\Pt_{22}-\Pt_{11}\} \,, \quad \{\Pt_{11} + \Pt_{22}\}\,.
\end{equation}
The proof then proceeds by checking, for each of the subalgebras $\mathfrak{h}$ of table \ref{tab:spin2contr}, which of these irreducible representations can be added to $\mathfrak{h}$, such that the enlarged $\mathfrak{h}$ forms a subalgebra that leads to a contraction with a non-abelian spin-3 part. Below we discuss the different contraction procedures. 

\begin{itemize}
\item Space-time contraction procedures: In this case we add irreducible representations (\ref{eq:irrepsJ}) to $\{\Jt, \Gt_a\}$.
  \begin{itemize}
  \item Adding $\Jt_a$, one finds from the commutator $[\Gt_a, \Jt_b]$ that one needs to add all $\Gt_{ab}$ (i.e.\ both irreducible representations $\{\Gt_{12},\Gt_{22}-\Gt_{11}\}$ and $\{\Gt_{11} + \Gt_{22}\}$) in order to obtain a subalgebra. One thus finds the subalgebra $\mathfrak{h} = \{\Jt, \Gt_a, \Jt_a, \Gt_{ab}\}$.
  \item Adding instead $\{\Gt_{12}, \Gt_{22} - \Gt_{11}\}$, one finds from the commutator of these two generators with $\Gt_a$ that one also needs to add $\Jt_a$ to obtain a subalgebra. From the commutator $[\Gt_a, \Jt_b]$, one then finds that one also needs to add all $\Gt_{ab}$. One thus again finds the subalgebra $\mathfrak{h} = \{\Jt, \Gt_a, \Jt_a, \Gt_{ab}\}$.
  \item Similarly, if one adds $\{\Gt_{11} + \Gt_{22}\}$ to $\{\Jt, \Gt_a\}$, one is again led to the subalgebra $\mathfrak{h} = \{\Jt, \Gt_a, \Jt_a, \Gt_{ab}\}$.
  \item It is not possible to add other spin-3 generators to $\{\Jt, \Gt_a, \Jt_a, \Gt_{ab}\}$, without ending up with a trivial contraction. Indeed, adding either $\{\Ht_a\}$, $\{\Pt_{12},\Pt_{22}-\Pt_{11}\}$ or $\{\Pt_{11} + \Pt_{22}\}$, one finds that requiring that one ends up with a subalgebra leads to a trivial contraction procedure.
\end{itemize}
Adding either $\{\Jt_a\}$, $\{\Gt_{12}, \Gt_{22}-\Gt_{11}\}$ or $\{\Gt_{11} + \Gt_{22}\}$, one thus only finds a non-trivial contraction procedure based on the subalgebra $\mathfrak{h} = \{\Jt, \Gt_a, \Jt_a, \Gt_{ab}\}$. The only other possibility for a space-time contraction procedure is thus obtained by adding $\{\Ht_a\}$, $\{\Pt_{12}, \Pt_{22}-\Pt_{11}\}$ or $\{\Pt_{11} + \Pt_{22}\}$ to $\{\Jt, \Gt_a\}$. Repeating the above reasoning with the replacement $\Jt_a \leftrightarrow \Ht_a$ and $\Gt_{ab} \leftrightarrow \Pt_{ab}$, one sees that the only non-trivial space-time contraction procedure based on this choice is given by the subalgebra $\{\Jt, \Gt_a, \Ht_a, \Pt_{ab}\}$.

We thus find that the only non-trivial space-time contraction procedures are the ones based on the subalgebras $\{\Jt, \Gt_a, \Jt_a, \Gt_{ab}\}$ and $\{\Jt, \Gt_a, \Ht_a, \Pt_{ab}\}$.
\item Speed-time contraction procedures: In this case we add irreducible representations (\ref{eq:irrepsJ}) to $\{\Jt, \Pt_a\}$. We can apply the same reasoning as for the space-time contraction procedures to find that the only non-trivial speed-time contraction procedures are the ones based on the subalgebras $\{\Jt, \Pt_a, \Jt_a, \Pt_{ab}\}$ and $\{\Jt, \Pt_a, \Ht_a, \Gt_{ab}\}$.
\item Speed-space contraction procedures: In this case we add irreducible representations (\ref{eq:irrepsJ}) to $\{\Jt, \Ht\}$.
  \begin{itemize}
  \item Adding $\Jt_a$, one finds from the commutator $[\Ht, \Jt_a]$ that one also needs to add $\Ht_a$ to obtain a subalgebra. Similarly, adding instead $\Ht_a$, one finds from the commutator $[\Ht, \Ht_a]$ that one needs to add $\Jt_a$ to obtain a subalgebra. Adding either $\Jt_a$ or $\Ht_a$ thus leads to a subalgebra $\{\Jt, \Ht, \Jt_a, \Ht_a\}$. This subalgebra is maximal in the sense that adding any other spin-3 generators leads to a trivial contraction procedure.
  \item Adding instead only $\Gt_{11} + \Gt_{22}$ or $\Pt_{11} + \Pt_{22}$ leads to a contraction with abelian spin-3 part, so this is excluded.
  \item Adding $\{\Gt_{12}, \Gt_{22} - \Gt_{11}\}$, one finds from their commutators with $\Ht$ that one also needs to add $\{\Pt_{12}, \Pt_{22} - \Pt_{11}\}$. Similarly, adding $\{\Pt_{12}, \Pt_{22} - \Pt_{11}\}$, one finds that one also needs to add $\{\Gt_{12}, \Gt_{22} - \Gt_{11}\}$. One thus finds a contraction procedure based on the subalgebra $\{\Jt, \Ht, \Gt_{12}, \Gt_{22} - \Gt_{11}, \Pt_{12}, \Pt_{22} - \Pt_{11}\}$. This subalgebra is not maximal as both $\Gt_{11} + \Gt_{22}$ and $\Pt_{11} + \Pt_{22}$ commute with it. One can thus obtain other contraction procedures based on the subalgebras $\{\Jt, \Ht, \Gt_{ab}, \Pt_{12}, \Pt_{22} - \Pt_{11}\}$, $\{\Jt, \Ht, \Gt_{12}, \Gt_{22} - \Gt_{11}, \Pt_{ab}\}$ and $\{\Jt, \Ht, \Gt_{ab}, \Pt_{ab}\}$.
  \end{itemize}
One can check that the above possibilities exhaust all possibilities for non-trivial speed-space contraction procedures. So, one finds that the only non-trivial speed-space contraction procedures are based on the subalgebras $\{\Jt, \Ht, \Jt_a, \Ht_a\}$, $\{\Jt, \Ht, \Gt_{12}, \Gt_{22} - \Gt_{11}, \Pt_{12}, \Pt_{22} - \Pt_{11}\}$, $\{\Jt, \Ht, \Gt_{ab}, \Pt_{12}, \Pt_{22} - \Pt_{11}\}$, $\{\Jt, \Ht, \Gt_{12}, \Gt_{22} - \Gt_{11}, \Pt_{ab}\}$ and $\{\Jt, \Ht, \Gt_{ab}, \Pt_{ab}\}$.
\item General contraction procedures: In this case we add irreducible representations (\ref{eq:irrepsJ}) to $\{\Jt\}$.
  \begin{itemize}
  \item Adding $\Jt_a$ leads to a subalgebra $\{\Jt, \Jt_a\}$ that obeys all requirements.
  \item Adding $\Ht_a$ similarly leads to a subalgebra $\{\Jt, \Ht_a\}$ that satisfies all requirements.
  \item Since $[\Jt_a, \Ht_b]$ generates $\Ht$, adding $\Jt_a$ and $\Ht_a$ simultaneously brings one back to the case of the speed-space contraction procedures that was already discussed above.
  \item Adding extra spin-3 generators to either $\{\Jt, \Jt_a\}$ or $\{\Jt, \Ht_a\}$ leads only to trivial contractions.
  \item Adding only $\Gt_{11} + \Gt_{22}$ or $\Pt_{11} + \Pt_{22}$ to $\{\Jt\}$ leads to a contraction with abelian spin-3 part, so this is excluded.
  \item Adding $\{\Gt_{12}, \Gt_{22} - \Gt_{11}\}$ or $\{\Pt_{12}, \Pt_{22} - \Pt_{11}\}$ to $\{\Jt\}$ leads to valid subalgebras. Adding them simultaneously, one finds that their commutator generates $\Ht$, bringing us back to the speed-space contractions already discussed above. Restricting oneself to general contractions, one thus finds two subalgebras $\{\Jt, \Gt_{12}, \Gt_{22} - \Gt_{11}\}$ and $\{\Jt, \Pt_{12}, \Pt_{22} - \Pt_{11}\}$. These are not maximal as both $\Gt_{11} + \Gt_{22}$ and $\Pt_{11} + \Pt_{22}$ commute with them. One can thus obtain other contraction procedures based on the subalgebras $\{\Jt, \Gt_{ab}\}$, $\{\Jt, \Pt_{ab}\}$, $\{\Jt, \Gt_{12}, \Gt_{22} - \Gt_{11}\}$, $\{\Jt, \Gt_{12}, \Gt_{22} - \Gt_{11}, \Pt_{11} + \Pt_{22}\}$, $\{\Jt, \Pt_{12}, \Pt_{22} - \Pt_{11}\}$ and $\{\Jt, \Pt_{12}, \Pt_{22} - \Pt_{11}, \Gt_{11} + \Gt_{22}\}$.
  \end{itemize}
The above reasoning exhausts all possible general contraction procedures, that are thus based on subalgebras $\{\Jt, \Jt_a\}$, $\{\Jt, \Ht_a\}$, $\{\Jt, \Gt_{ab}\}$, $\{\Jt, \Pt_{ab}\}$, $\{\Jt, \Gt_{12}, \Gt_{22} - \Gt_{11}\}$, $\{\Jt, \Gt_{12}, \Gt_{22} - \Gt_{11}, \Pt_{11} + \Pt_{22}\}$, $\{\Jt, \Pt_{12}, \Pt_{22} - \Pt_{11}\}$ and $\{\Jt, \Pt_{12}, \Pt_{22} - \Pt_{11}, \Gt_{11} + \Gt_{22}\}$.
\end{itemize}

\section{Explicit Commutation Relations of the Democratic Spin-3 Algebras}
\label{app:all-expl-contr}

This appendix contains tables with all the commutation relations of the spin-3 algebras that can be obtained via sequential application of the `democratic' IW contraction procedures. We start each  table with the spin-2 commutation relations, then proceed with the mixed spin commutation relations and conclude with the spin-3 commutation relations. The table caption contains information about what type of higher spin version we are dealing with (e.g.~higher spin version of Poincar\'e, Galilei or Carroll). Under the heading `Contraction \#', we have indicated one possibility of obtaining the corresponding algebra as a sequential application of IW contraction procedures. The numbers in this heading refer to the contraction procedures of table \ref{tab:contr}.

For layout reasons the tables start on the next page.

\newpage

\begin{table}[H]
  \centering
$
\begin{array}{l r r r r r}
\toprule %
                                               & \mathfrak{hs}_{3}\mathfrak{(A)dS}_{\parmp}          & \mathfrak{hs}_{3}\mathfrak{poi1}                  & \mathfrak{hs}_{3}\mathfrak{poi2}              \\
  \text{Contraction \#}                        &                                                     & 1                                                 & 2                                                   \\ \midrule
  \left[\,\Jt  \comma \Gt_{a} \,\right]        & \epsilon_{am}  \Gt_{m}                              & \epsilon_{am}  \Gt_{m}                            & \epsilon_{am}  \Gt_{m}                              \\
  \left[\, \Jt\comma \Ht \,\right]             & 0                                                   & 0                                                 & 0                                                   \\
  \left[\, \Jt \comma \Pt_{a} \,\right]        & \epsilon_{am}  \Pt_{m}                              & \epsilon_{am}  \Pt_{m}                            & \epsilon_{am}  \Pt_{m}                              \\ 
  \left[\,\Gt_{a}  \comma \Gt_{b} \,\right]    & - \epsilon_{ab}  \Jt                                & - \epsilon_{ab}  \Jt                              & - \epsilon_{ab}  \Jt                                \\
  \left[\, \Gt_{a} \comma \Ht \,\right]        & -\epsilon_{am}  \Pt_{m}                             & -\epsilon_{am}  \Pt_{m}                           & -\epsilon_{am}  \Pt_{m}                             \\
  \left[\, \Gt_{a} \comma \Pt_{b} \,\right]    & -\epsilon_{ab} \Ht                                  & -\epsilon_{ab} \Ht                                & -\epsilon_{ab} \Ht                                  \\
\end{array}
$
\caption{Higher spin versions of the (A)dS and Poincar\'e algebra. The upper sign is for AdS (and contractions thereof) and the lower sign for dS (and contractions thereof).}
\label{tab:adspoin}
\end{table}

\begin{table}[H]
  \centering
$
\begin{array}{l r r r r r}
\toprule %
                                               & \mathfrak{hs}_{3}\mathfrak{nh1}                          & \mathfrak{hs}_{3}\mathfrak{nh2}                    \\
  \text{Contraction \#}                           & 3                                                   & 4                                             \\ \midrule
  \left[\,\Jt  \comma \Gt_{a} \,\right]        & \epsilon_{am}  \Gt_{m}                              & \epsilon_{am}  \Gt_{m}                        \\
  \left[\, \Jt\comma \Ht \,\right]             & 0                                                   & 0                                             \\
  \left[\, \Jt \comma \Pt_{a} \,\right]        & \epsilon_{am}  \Pt_{m}                              & \epsilon_{am}  \Pt_{m}                        \\ 
  \left[\,\Gt_{a}  \comma \Gt_{b} \,\right]    & 0                                                   & 0                                             \\
  \left[\, \Gt_{a} \comma \Ht \,\right]        & -\epsilon_{am}  \Pt_{m}                             & -\epsilon_{am}  \Pt_{m}                       \\
  \left[\, \Gt_{a} \comma \Pt_{b} \,\right]    & 0                                                   & 0                                             \\
\end{array}
$
\caption{Higher spin versions of the Newton--Hooke algebra. The upper sign is for contractions of AdS and the lower sign for contractions of dS.}
\label{tab:nh}
\end{table}

\begin{table}[H]
  \centering
$
\begin{array}{l r r r r r}
\toprule %
                                               & \mathfrak{hs}_{3}\mathfrak{ppoi1}                        & \mathfrak{hs}_{3}\mathfrak{ppoi2}                        \\
  \text{Contraction \#}                        & 5                                                   & 6                                                   \\ \midrule
  \left[\,\Jt  \comma \Gt_{a} \,\right]        & \epsilon_{am}  \Gt_{m}                              & \epsilon_{am}  \Gt_{m}                              \\
  \left[\, \Jt\comma \Ht \,\right]             & 0                                                   & 0                                                   \\
  \left[\, \Jt \comma \Pt_{a} \,\right]        & \epsilon_{am}  \Pt_{m}                              & \epsilon_{am}  \Pt_{m}                              \\ 
  \left[\,\Gt_{a}  \comma \Gt_{b} \,\right]    & 0                                                   & 0                                                   \\
  \left[\, \Gt_{a} \comma \Ht \,\right]        & 0                                                   & 0                                                   \\
  \left[\, \Gt_{a} \comma \Pt_{b} \,\right]    & -\epsilon_{ab} \Ht                                  & -\epsilon_{ab} \Ht                                  \\
\end{array}
$
\caption{Higher spin versions of Para-Poincar\'e algebra. The upper sign is for contractions of AdS and the lower sign for contractions of dS.}
\label{tab:poins}
\end{table}

\begin{table}[H]
  \centering
$
\begin{array}{l r r r r}
\toprule %
                                               & \mathfrak{hs}_{3}\mathfrak{car1}                             & \mathfrak{hs}_{3}\mathfrak{car2}                             & \mathfrak{hs}_{3}\mathfrak{car3}                         & \mathfrak{hs}_{3}\mathfrak{car4}                         \\
  \text{Contraction \#}                        & 1,5                                               & 1,6                                               & 5,2                                                 & 6,2                                                 \\ \midrule
  \left[\,\Jt  \comma \Gt_{a} \,\right]        & \epsilon_{am}  \Gt_{m}                            & \epsilon_{am}  \Gt_{m}                            & \epsilon_{am}  \Gt_{m}                              & \epsilon_{am}  \Gt_{m}                              \\
  \left[\, \Jt\comma \Ht \,\right]             & 0                                                 & 0                                                 & 0                                                   & 0                                                   \\
  \left[\, \Jt \comma \Pt_{a} \,\right]        & \epsilon_{am}  \Pt_{m}                            & \epsilon_{am}  \Pt_{m}                            & \epsilon_{am}  \Pt_{m}                              & \epsilon_{am}  \Pt_{m}                              \\ 
  \left[\,\Gt_{a}  \comma \Gt_{b} \,\right]    & 0                                                 & 0                                                 & 0                                                   & 0                                                   \\
\end{array}
$
\caption{Higher spin versions of the Carroll algebra. The upper sign is for contractions of AdS and the lower sign for contractions of dS.}
\label{tab:hscar}
\end{table}

\begin{table}[H]
  \centering
$
\begin{array}{l r r r r r}
\toprule %
                                               & \mathfrak{hs}_{3}\mathfrak{gal1}                             & \mathfrak{hs}_{3}\mathfrak{gal2}                      & \mathfrak{hs}_{3}\mathfrak{gal3}                         & \mathfrak{hs}_{3}\mathfrak{gal4}                   \\
  \text{Contraction \#}                        & 1,3                                               & 1,4                                        & 3,2                                                 & 4,2                                           \\ \midrule
  \left[\,\Jt  \comma \Gt_{a} \,\right]        & \epsilon_{am}  \Gt_{m}                            & \epsilon_{am}  \Gt_{m}                     & \epsilon_{am}  \Gt_{m}                              & \epsilon_{am}  \Gt_{m}                        \\
  \left[\, \Jt\comma \Ht \,\right]             & 0                                                 & 0                                          & 0                                                   & 0                                             \\
  \left[\, \Jt \comma \Pt_{a} \,\right]        & \epsilon_{am}  \Pt_{m}                            & \epsilon_{am}  \Pt_{m}                     & \epsilon_{am}  \Pt_{m}                              & \epsilon_{am}  \Pt_{m}                        \\ 
  \left[\,\Gt_{a}  \comma \Gt_{b} \,\right]    & 0                                                 & 0                                          & 0                                                   & 0                                             \\
\end{array}
$
\caption{Higher spin versions of the Galilei algebra. The upper sign is for contractions of AdS and the lower sign for contractions of dS.}
\label{tab:hsgal}
\end{table}

\begin{table}[H]
  \centering
$
\begin{array}{l r r r r r}
\toprule %
                                               & \mathfrak{hs}_{3}\mathfrak{pgal1}             & \mathfrak{hs}_{3}\mathfrak{pgal2}             & \mathfrak{hs}_{3}\mathfrak{pgal3} & \mathfrak{hs}_{3}\mathfrak{pgal4} \\
  \text{Contraction \#}                        & 3,5                                                 & 3,6                                                 & 4,5                                     & 4,6                                     \\ \midrule
  \left[\,\Jt  \comma \Gt_{a} \,\right]        & \epsilon_{am}  \Gt_{m}                              & \epsilon_{am}  \Gt_{m}                              & \epsilon_{am}  \Gt_{m}                  & \epsilon_{am}  \Gt_{m}                  \\
  \left[\, \Jt\comma \Ht \,\right]             & 0                                                   & 0                                                   & 0                                       & 0                                       \\
  \left[\, \Jt \comma \Pt_{a} \,\right]        & \epsilon_{am}  \Pt_{m}                              & \epsilon_{am}  \Pt_{m}                              & \epsilon_{am}  \Pt_{m}                  & \epsilon_{am}  \Pt_{m}                  \\ 
  \left[\,\Gt_{a}  \comma \Gt_{b} \,\right]    & 0                                                   & 0                                                   & 0                                       & 0                                       \\
  \left[\, \Gt_{a} \comma \Ht \,\right]        & 0                                                   & 0                                                   & 0                                       & 0                                       \\
\end{array}
$
\caption{Higher spin versions of the  Para-Galilei algebra. The upper sign is for contractions of AdS and the lower sign for contractions of dS.}
\label{tab:pgal}
\end{table}

\begin{table}[H]
  \centering
$
\begin{array}{l r r r r r}
\toprule %
                                               & \mathfrak{hs}_{3}\mathfrak{st1}                              & \mathfrak{hs}_{3}\mathfrak{st2}                       & \mathfrak{hs}_{3}\mathfrak{st3}                          & \mathfrak{hs}_{3}\mathfrak{st4}                    \\
  \text{Contraction \#}                        & 1,3,5=7                                           & 1,4,6=8                                    & 2,3,6=9                                             & 2,4,5=10                                      \\ \midrule
  \left[\,\Jt  \comma \Gt_{a} \,\right]        & \epsilon_{am}  \Gt_{m}                            & \epsilon_{am}  \Gt_{m}                     & \epsilon_{am}  \Gt_{m}                              & \epsilon_{am}  \Gt_{m}                        \\
  \left[\, \Jt\comma \Ht \,\right]             & 0                                                 & 0                                          & 0                                                   & 0                                             \\
  \left[\, \Jt \comma \Pt_{a} \,\right]        & 0                                                 & 0                                          & 0                                                   & 0                                             \\ 
  \left[\,\Gt_{a}  \comma \Gt_{b} \,\right]    & 0                                                 & 0                                          & 0                                                   & 0                                             \\
\end{array}
$
\caption{Higher spin versions of the static algebra. The upper sign is for contractions of AdS and the lower sign for contractions of dS.}
\label{tab:hsstat}
\end{table}

\begin{table}[H]
  \centering
$
\begin{array}{l r r r r r}
\toprule %
                                               & \mathfrak{hs}_{3}\mathfrak{st5}                              & \mathfrak{hs}_{3}\mathfrak{st8}                    & \mathfrak{hs}_{3}\mathfrak{st6}                       & \mathfrak{hs}_{3}\mathfrak{st7}                          \\
  \text{Contraction \#}                        & 3,1,6                                             & 4,2,6                                         & 1,4,5                                      & 3,2,5                                               \\ \midrule
  \left[\,\Jt  \comma \Gt_{a} \,\right]        & \epsilon_{am}  \Gt_{m}                            & \epsilon_{am}  \Gt_{m}                        & \epsilon_{am}  \Gt_{m}                     & \epsilon_{am}  \Gt_{m}                              \\
  \left[\, \Jt\comma \Ht \,\right]             & 0                                                 & 0                                             & 0                                          & 0                                                   \\
  \left[\, \Jt \comma \Pt_{a} \,\right]        & \epsilon_{am}  \Pt_{m}                            & \epsilon_{am}  \Pt_{m}                        & \epsilon_{am}  \Pt_{m}                     & \epsilon_{am}  \Pt_{m}                              \\ 
  \left[\,\Gt_{a}  \comma \Gt_{b} \,\right]    & 0                                                 & 0                                             & 0                                          & 0                                                   \\
\end{array}
$
\caption{Higher spin versions of the static algebra which can not be directly contracted. The upper sign is for contractions of AdS and the lower sign for contractions of dS.}
\label{tab:hsstat2}
\end{table}


\providecommand{\href}[2]{#2}\begingroup\raggedright\endgroup

\end{document}